\input epsf
\documentstyle{amsppt}
\pagewidth{5.4truein}\hcorrection{0.55in}
\pageheight{7.5truein}\vcorrection{0.75in}
\TagsOnRight
\NoRunningHeads
\catcode`\@=11
\def\logo@{}
\footline={\ifnum\pageno>1 \hfil\folio\hfil\else\hfil\fi}
\topmatter
\title A triangular gap of size two in a sea of dimers \\
in a $90^\circ$ angle with mixed boundary conditions,\\
and a heat flow conjecture for the general case
\endtitle
\vskip-0.2in
\author Mihai Ciucu\endauthor
\thanks Research supported in part by NSF grant DMS-1101670.
\endthanks
\affil
  Department of Mathematics, Indiana University\\
  Bloomington, Indiana 47405
\endaffil
\abstract
We consider a triangular gap of side two in a $90^\circ$ angle on the triangular lattice with mixed boundary conditions: a constrained, zig-zag boundary along one side, and a free lattice line boundary along the other. We study the interaction of the gap with the corner as the rest of the angle is completely filled with lozenges. We show that the resulting correlation is governed by the product of the distances between the gap and its three images in the sides of the angle. 
The image in the side with constrained boundary has the same orientation as the original gap, while the image in the side with free boundary has the opposite orientation. This, together with the parallel between the correlation of gaps in dimer packings and electrostatics we developed in previous work, provides evidence for a unified way of understanding the interaction of gaps with the bounary under mixed boundary conditions, which we phrase as a conjecture. While the electrostatic interpretation is equivalent to a steady state heat flow interpretation in the bulk, it turns out that the latter view is more natural in the context of the interaction of the gaps with the boundary.

The starting point for our analysis is an exact formula we prove for the number of lozenge tilings of certain trapezoidal regions with mixed boundary conditions, which is equivalent to a new, multi-parameter generalization of a classical plane partition enumeration problem (that of enumerating symmetric, self-complementary plane partitions).
\endabstract
\endtopmatter

\document

\def\mysec#1{\bigskip\centerline{\bf #1}\message{ * }\nopagebreak\bigskip\par}

\def\myref#1{\item"{[{\bf #1}]}"}

\def\pf{{\it Proof.\ }}

\def\epf{\hfill{$\square$}\smallpagebreak}

\def\cite#1{\relaxnext@
  \def\nextiii@##1,##2\end@{[{\bf##1},\,##2]}%
  \in@,{#1}\ifin@\def\next{\nextiii@#1\end@}\else
  \def\next{[{\bf#1}]}\fi\next}
\def\proclaimheadfont@{\smc}

\def\pf{{\it Proof.\ }}

\define\Z{{\Bbb Z}}

\define\M{\operatorname{M}}
\define\de{\operatorname{d}}
\define\q{\operatorname{q}}
\define\Pf{\operatorname{Pf}}
\define\sgn{\operatorname{sgn}}
\define\al{\alpha}
\define\be{\beta}


\define\twoline#1#2{\line{\hfill{\smc #1}\hfill{\smc #2}\hfill}}
\define\twolinetwo#1#2{\line{{\smc #1}\hfill{\smc #2}}}
\define\twolinethree#1#2{\line{\phantom{poco}{\smc #1}\hfill{\smc #2}\phantom{poco}}}

\def\mypic#1{\epsffile{#1}}



\define\And{1}
\define\ri{2}
\define\sc{3}
\define\ec{4}
\define\ef{5}
\define\ov{6}
\define\free{7}
\define\gd{8}
\define\anglep{9}
\define\aanglep{10}
\define\fakt{11}
\define\DT{12}
\define\Feyntwo{13}
\define\FS{14}
\define\GV{15}
\define\Kuo{16}
\define\Kup{17}
\define\Lind{18}
\define\Proc{19}
\define\Schur{20}
\define\Sta{21}
\define\Ste{22}


\define\eba{2.1}
\define\ebb{2.2}
\define\ebc{2.3}
\define\ebd{2.4}
\define\ebe{2.5}

\define\eca{3.1}
\define\ecb{3.2}
\define\ecc{3.3}
\define\ecca{3.4}
\define\eccab{3.5}
\define\eccb{3.6}
\define\eccc{3.7}
\define\eccd{3.8}
\define\ecd{3.9}
\define\ece{3.10}
\define\ecf{3.11}
\define\ecg{3.12}
\define\ech{3.13}
\define\eci{3.14}
\define\ecj{3.15}
\define\eck{3.16}
\define\ecl{3.17}
\define\ecm{3.18}
\define\ecn{3.19}
\define\eco{3.20}

\define\eda{4.1}
\define\edb{4.2}
\define\edc{4.3}
\define\edd{4.4}
\define\ede{4.5}

\define\eea{5.1}
\define\eeb{5.2}
\define\eec{5.3}
\define\eed{5.4}
\define\eee{5.5}
\define\eef{5.6}
\define\eeg{5.7}
\define\eeh{5.8}
\define\eei{5.9}

\define\efa{6.1}
\define\efb{6.2}
\define\efc{6.3}
\define\efd{6.4}
\define\efe{6.5}

\define\ega{7.1}
\define\egb{7.2}
\define\egc{7.3}
\define\egd{7.4}
\define\ege{7.5}
\define\egf{7.6}
\define\egg{7.7}

\define\eia{8.1}

\define\eja{9.1}


\define\tba{2.1}
\define\tbb{2.2}

\define\tca{3.1}
\define\tcb{3.2}
\define\tcc{3.3}
\define\tcd{3.4}

\define\tda{4.1}

\define\tea{5.1}

\define\tfa{6.1}

\define\tia{8.1}


\define\fbc{2.3}
\define\fbd{2.4}
\define\fbe{2.5}

\define\fcb{3.1}
\define\fcc{3.2}
\define\fcd{3.3}
\define\fce{3.4}
\define\fcf{3.5}
\define\fcg{3.6}

\define\fea{5.1}
\define\feb{5.2}
\define\fec{5.3}
\define\fed{5.4}
\define\fee{5.5}

\define\fia{8.1}
\define\fib{8.2}

\vskip-0.05in
\mysec{1. Introduction}

The study of the correlation of gaps in a sea of dimers was initiated five decades ago by Fisher and Stephenson in their paper \cite{\FS}, where, based on precise numerical calculations, they conjectured that the correlation of two monomers in an otherwise closely packed dimer system on the square grid is rotationally invariant in the scaling limit. 

In a previous series of articles (see \cite{\ri}\cite{\sc}\cite{\ec}\cite{\gd}), we extended the problem of Fisher and Stephenson to the situation when one is allowed to have any finite number of gaps, each of an arbitrary size, and we showed that a close parallel to electrostatics emerges: As the distances between the gaps approach infinity, their correlation is given by the exponential of the negative of the electrostatic energy of a two dimensional system of charges that correspond to the gaps in a natural way. This parallel to electrostatics was then extended in \cite{\ef} and \cite{\ov}, where we showed that the discrete field of the average tile orientations approaches, in the scaling limit, the electric field. 

Further aspects of this analogy concern the behavior of the correlation of gaps near the boundary of lattice regions enclosing them, which in previous work was shown to be analogous to the behavior of electrical charges near conductors\footnote{ We will see, however, that another physical set-up, namely the steady state heat flow problem (to which the electrostatics of point charges is equivalent in the bulk) provides an even closer parallel to the interaction of gaps in dimer packings in the presence of a boundary. This is explained in Section 8.}. The case of a half-plane with constrained boundary was treated in \cite{\sc}, where it was shown that the asymptotics of the correlation of gaps on the triangular lattice near a constrained zig-zag boundary is given by a variant of the method of images from electrostatics, in which the image charges have the same signs as the original ones.
 The case of a $60^\circ$ angle was discussed in \cite{\anglep}, where we proved that the interaction of a triangular gap of side two with such an angle on the triangular lattice, having constrained zig-zag boundaries along both sides, is also given by the above mentioned variant of the method of images (which, in this situation, creates five images of the original gap). The only example in the previous literature involving a gap near a free boundary in the one presented in \cite{\free}, which concerns a triangular gap of side two on the triangular lattice near a free lattice line boundary; the asymptotics of the correlation is governed in that situation by the actual method of images from physics (i.e. image charges have opposite signs).

In this paper we study the interaction of a gap in an angle with mixed boundary conditions, a situation that is not covered in the previous literature. More precisely, we consider a triangular gap of side two inside a $90^\circ$ angle on the triangular lattice with constrained zig-zag boundary along one side, and free lattice line boundary along the other, and we determine the asymptotics of the correlation of the gap as it recedes along an arbitrary ray through the vertex of the angle (see Theorem {\tba}). We find that the answer is given by a version of the method of images involving two different sign rules --- opposite signs for the images in the free boundary, and same sign for the images in the constrained boundary.

This adds an essential new entry to the small collection of such examples worked out in the literature. Beyond the proof of Theorem 2.1, the main point of the current paper is that these five separate special circumstances, and the results they lead to, can be phrased in a unified way in terms of the 2D steady state heat flow problem in a uniform block of material having a finite number of point heat sources and sinks in its interior, and mixed boundary conditions --- Neumann boundary conditions (perfectly insulating boundary) along the portions corresponding to constrained zig-zag boundaries, and Dirichlet boundary conditions (boundary kept at a common constant temperature) along the portions corresponding to free lattice line boundaries. Based on this evidence, we conjecture that in the general case the interaction of gaps with these kinds of boundaries is given by the corresponding steady state heat flow problem (see Conjecture {\tia}). We detail these considerations in Section 8.


We note that, from the point of view of the literature on plane partitions and their symmetry classes (see for instance \cite{\Sta}, \cite{\And}, \cite{\Ste} and \cite{\Kup}), Corollary {\tcd} of this paper --- which is equivalent to Proposition {\tca}, the starting point of our asymptotic analysis of the interaction of the gap with the $90^\circ$ angle --- represents a new, multi-parameter generalization of the symmetric, self-complementary case, first solved by Proctor \cite{\Proc}.


%
%
%
%

\mysec{2. Statement of the main result and physical interpretation}

Let $n\geq2$, $x\geq0$ and $y\geq -1$ be integers. Consider the pentagonal region illustrated by Figure~{\tba}, where the western side has length $x$, the southeastern side has length\footnote{ The reason we denote the length of the southeastern side by $y+1$ rather than $y$ is that this way the formula in Proposition {\tca} is slightly simpler.} $y+1$, and the bottom side follows a zig-zag lattice path of length $2n$. In addition, we let the boundary along the eastern side be {\it free}, meaning that when considering a lozenge tiling of this region\footnote{ A lozenge tiling of a lattice region on the triangular lattice is a covering of that region by rhombi consisting of the union of two unit triagles sharing an edge; such unit rhombi are called lozenges.}, lozenges are allowed to protrude out halfway along this boundary (an example of a lozenge tiling of a related region that has such a free boundary is pictured in Figure~{\fcc}).  Denote the resulting region by $D_{n,x,y}$.

For  positive integers $\al$ and $\be$, we define $D_{n,x,y}(\al,\be)$ to be the region obtained from $D_{n,x,y}$ by removing from it the left-pointing lattice triangle of side-length 2 whose top side's center has coordinates $(\al\sqrt{3},\be)$ in the rectangular system of coordinates indicated in Figure {\tbb}.

\topinsert
\twolinetwo{\mypic{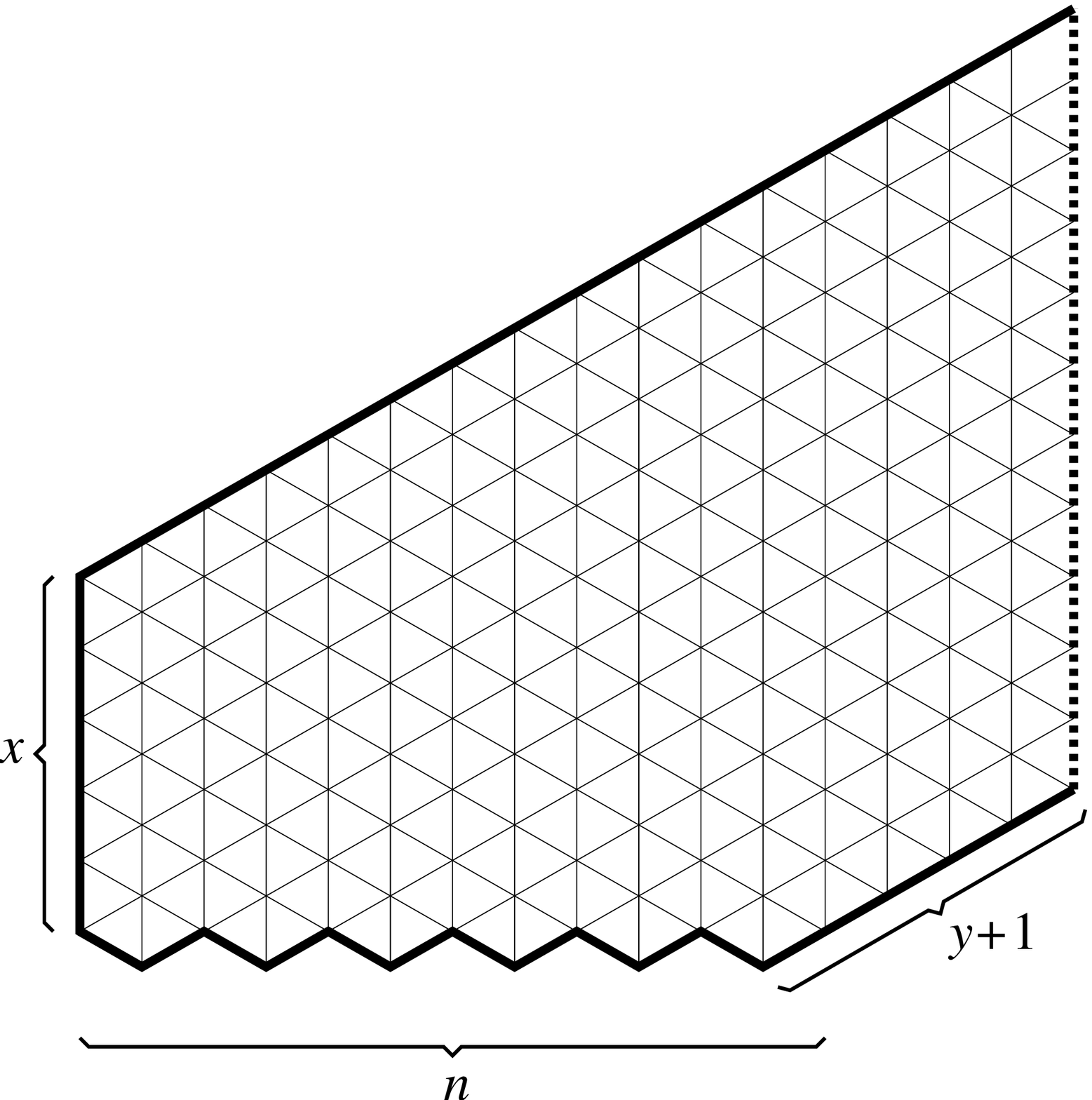}}{\mypic{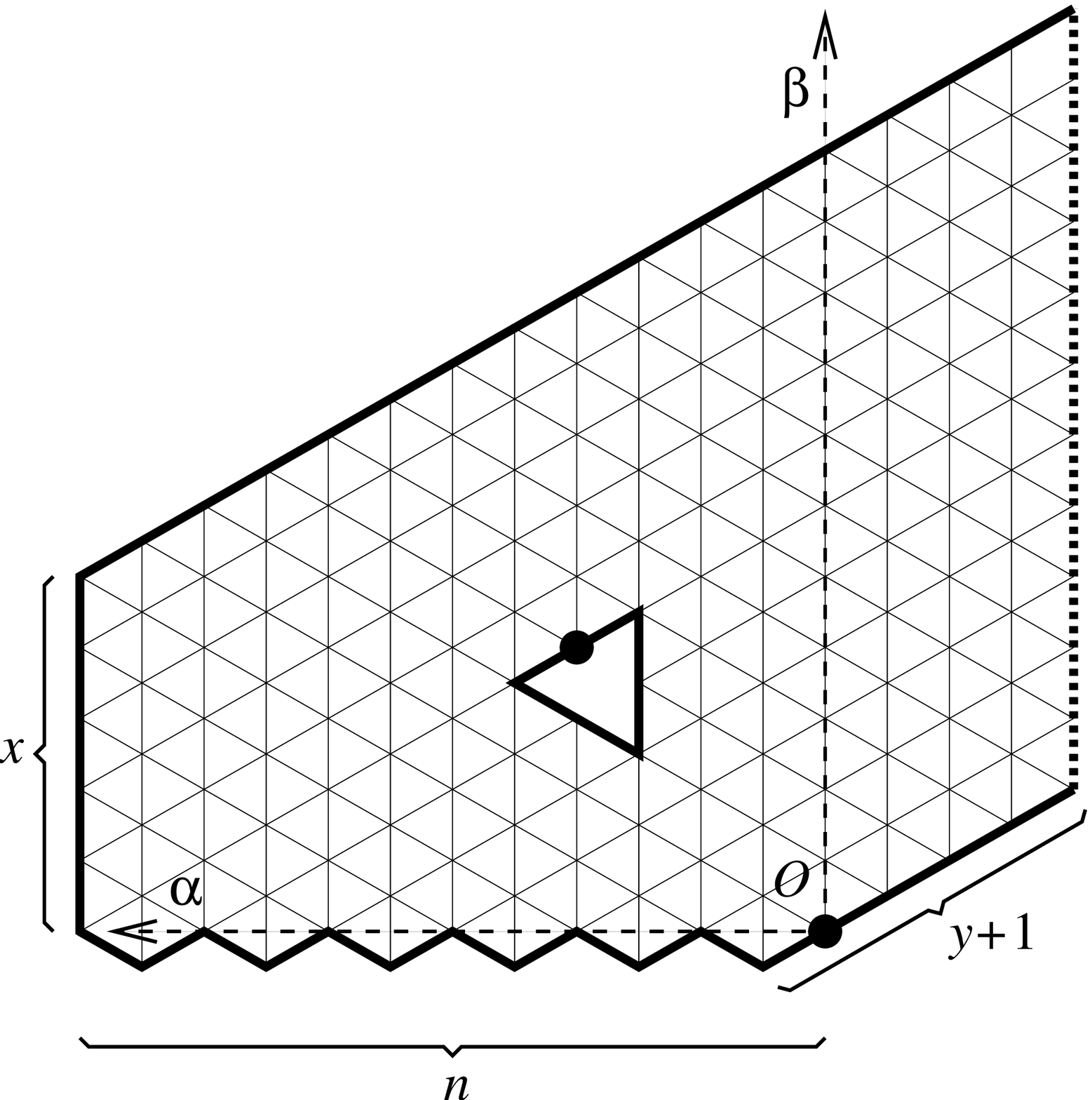}}
\twoline{Figure~{\tba}. {\rm $D_{6,5,4}$.}}
{Figure~{\tbb}. {\rm  $D_{6,5,4}(2,4)$.}}
\endinsert


For $x=n$, $y=0$ and fixed $\al$ and $\be$, as $n$ grows to infinity the gap is effectively in an infinite angular region, one of whose sides is a {\it constrained} zig-zag lattice path, and the other a {\it free} lattice line, meeting at a $90^\circ$ angle. Define the correlation $\omega_c(\al,\be)$ of the gap with the corner of this angle by
$$
\omega_c(\al,\be):=\lim_{n\to\infty}\frac{\M_f(D_{n,n,0}(\al,\be))}{\M_f(D_{n,n,0}(1,1))},
\tag\eba
$$
where, for a lattice region $D$ on the triangular lattice part of whose boundary is free, $\M_f(D)$ denotes the number of lozenge tilings of $D$ (the subscript of $\omega$  indicates that the correlation feels the interaction with the corner of the angle, as $R$ and $v$ are fixed; the subscript of $\M$ indicates that the tilings we are counting are allowed to protrude out through the free portion of the boundary). 
In the special case $R=3$, $v=4$ and $n=10$, the regions at the numerator and denominator on the right hand side of (\eba) are shown in Figures {\fbc} and~{\fbd}, respectively.

\topinsert
\twolinetwo{\mypic{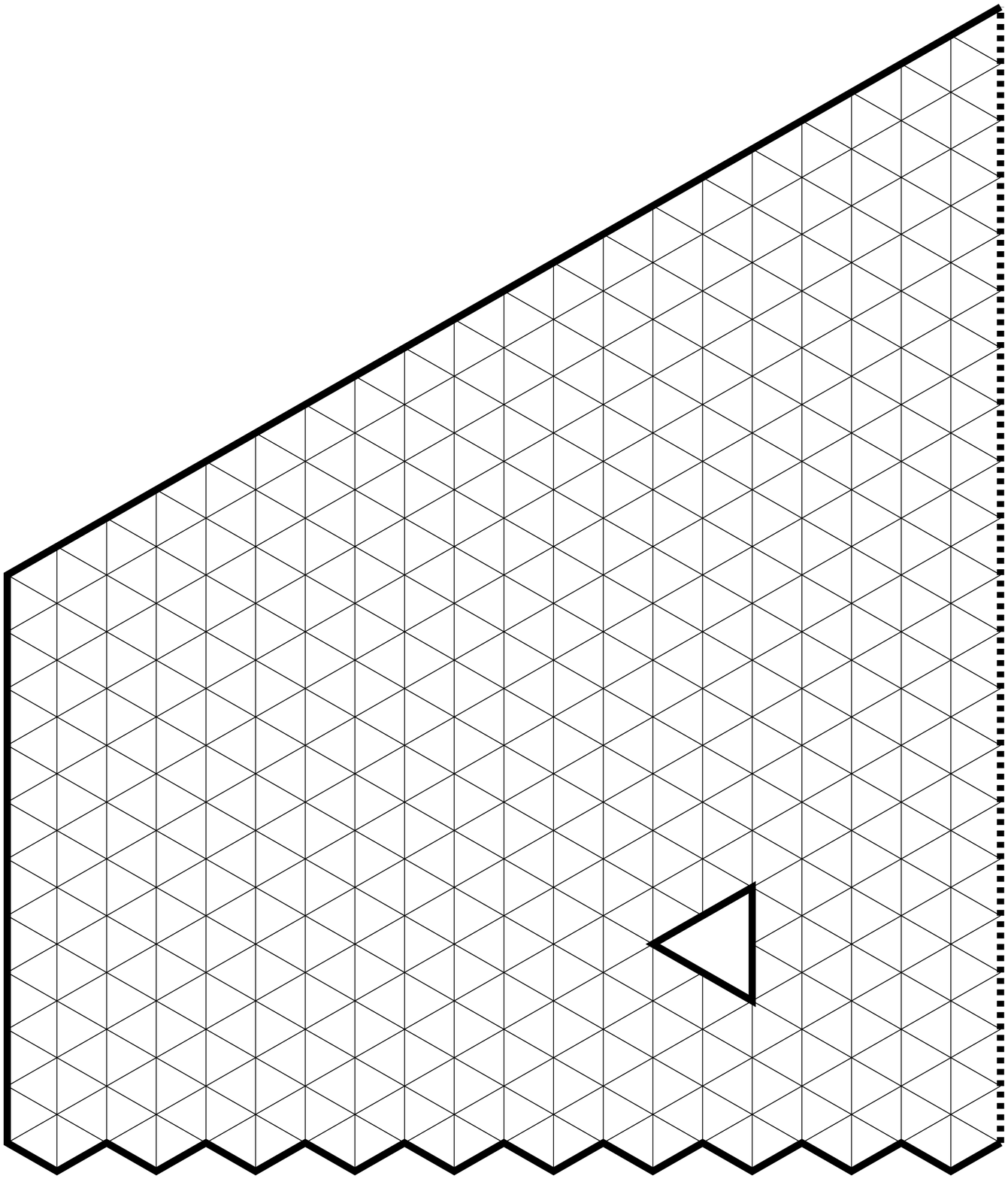}}{\ \ \ \ \ \ \mypic{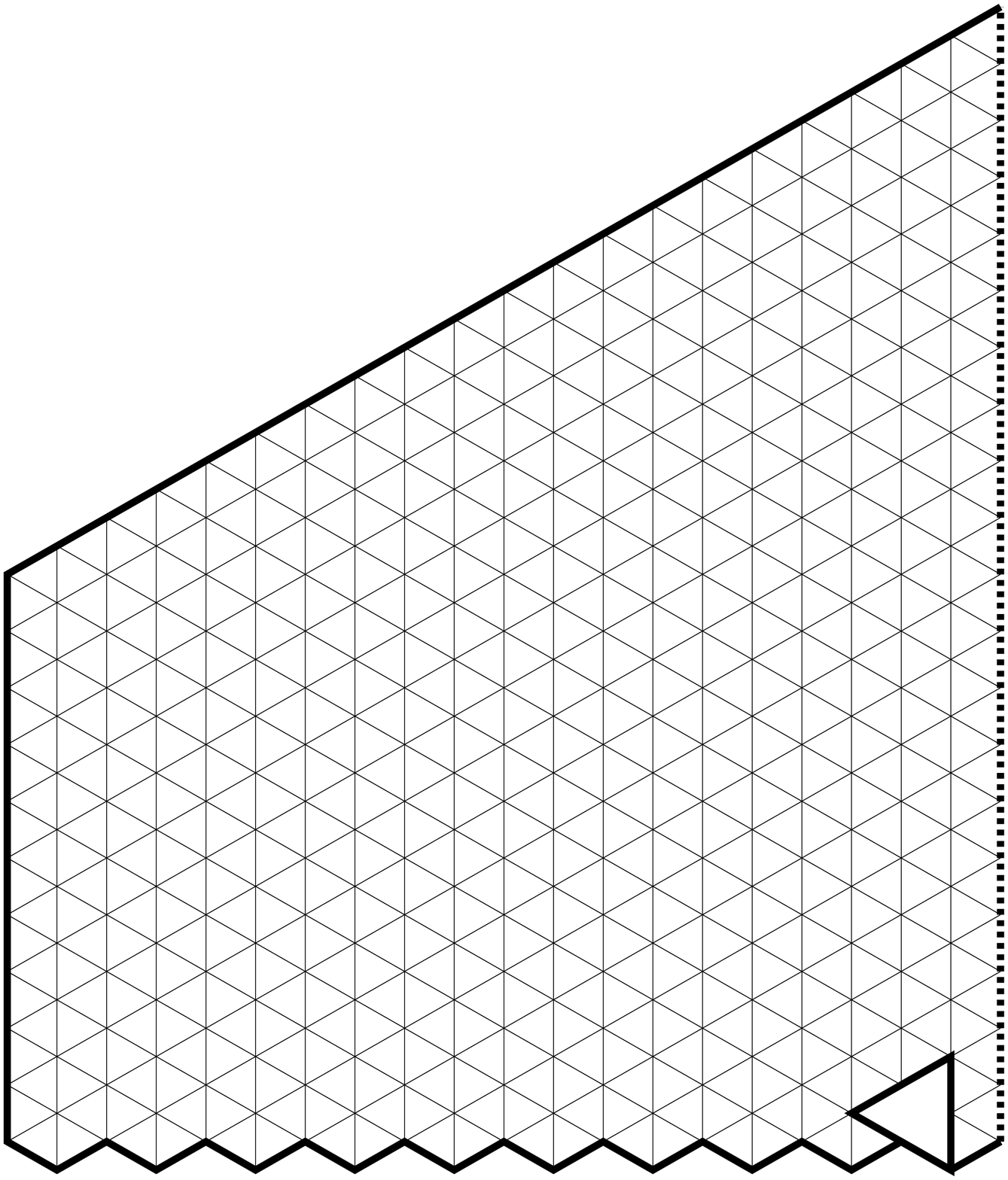}}
\twoline{Figure~{\fbc}. {\rm $D_{10,10,0}(3,4)$.}}
{Figure~{\fbd}. {\rm $D_{10,10,0}(1,1)$.}}
\endinsert



The main result of this paper is the following.

Let us denote our triangular gap of side two inside the $90^\circ$ angle by $O_1$, and let $\ell_1$ and $\ell_2$ be the straight lines supporting the horizontal zig-zag lattice path and the vertical free boundary, respectively (they are indicated by dashed lines in Figure {\fbe}). Let $O_2$ and $O_3$ be the mirror images of $O_1$ in $\ell_1$ and $\ell_2$, respectively. 
Then the mirror image of $O_2$ in $\ell_1$ is the same as the mirror image of $O_3$ in $\ell_2$; denote it by $O_4$. (Note that $\{O_2,O_3,O_4\}$ is the set of all images $O_1$ would see if the sides of the angle were mirrors.)

\proclaim{Theorem \tba} Let $q$ be a fixed positive rational number. As $\al$ and $\be$ approach infinity so that $\al=q\be$, we have
$$
\spreadlines{4\jot}
\align
\omega_c(\al,\be)
\sim
\frac{16}{3\pi Rq \sqrt{q^2+\frac13}}
\sim
\frac{32}{\pi}\,
\sqrt{\frac{\de(O_1,O_2)\de(O_3,O_4)}{\de(O_1,O_3)\de(O_1,O_4)\de(O_2,O_3)\de(O_2,O_4)}},
\tag\ebb
\endalign
$$
where $\de$ is the Euclidean distance.

\endproclaim

\topinsert
\centerline{\mypic{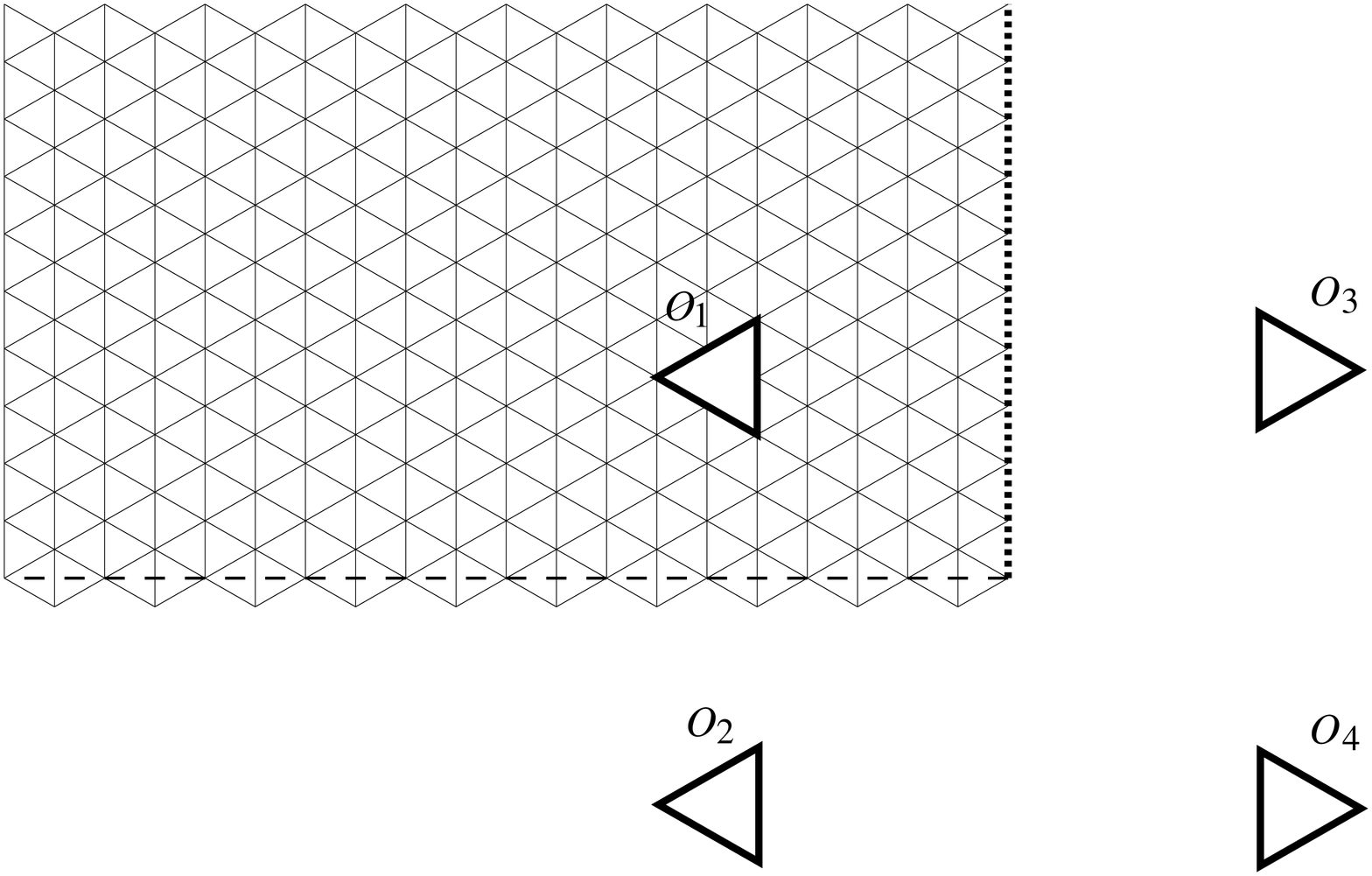}}
\medskip
\centerline{{\smc Figure~{\fbe}. {\rm  The gap and its three images for $\al=3$, $\be=4$.}}}
\endinsert

\flushpar
{\smc Remark 1.} In \cite{\ec} we showed that if $O_1,\dotsc,O_n$ are unions of collinear triangles of side two (which can point left or right, but are of the same kind within each $O_i$), then, for large separations between the $O_i$'s, the asymptotics of their correlation in the bulk is given by
$$
\omega(O_1,\dotsc,O_n)\sim c \prod_{1\leq i<j \leq n} \de(O_i,O_j)^{\frac12 \q(O_i)\q(O_j)},
\tag\ebc
$$
where $\q(O)$ denotes the charge of the gap $O$, defined to be the number of right-pointing unit triangles in $O$ minus the number of left-pointing unit triangles in $O$, and the multiplicative constant $c$ depends only on the structure of the individual gaps, and not on their relative position.

Note that, since our gap $O_1$ and its doubly reflected image $O_4$ have charge equal to $-2$, while the reflections $O_2$ and $O_3$ have charge $+2$, the asymptotics (\ebb) of the correlation of the gap in our $90^\circ$ angle with mixed boundary conditions can be written as
$$
\frac{32}{\pi}\,
\root{4}\of{\prod_{1\leq i<j\leq 4} \de(O_i,O_j)^{\frac12 \q(O_i)\q(O_j)}}.
\tag\ebd
$$
Thus, using (\ebc), one can rewrite the statement of Theorem {\tba} as
$$
\omega_c(O_1)\sim c'\root{4}\of{\omega(O_1,\dotsc,O_4)},
\tag\ebe
$$
where $c'$ is some explicit numerical constant. Thus the correlation at the corner can be expressed in terms of the correlation in the bulk of the gap with its images, much like in the method of images of electrostatics, when the electric field created by a charge near a conductor can be found by replacing the conductor with a suitable system of image charges. 

In fact, an even closer physical analog --- indeed a perfect parallel --- turns out to be the steady state heat flow problem in a uniform block of material, with a finite number of point heat sources and sinks (corresponding to the gaps), and two kinds of boundary conditions: Neumann boundary conditions (namely, perfectly insulating boundary) along the portions corresponding to constrained zig-zag boundaries, and Dirichlet boundary conditions (namely, boundary kept at a common constant temperature) along the portions corresponding to free lattice line boundaries.

This interpretation turns out to fit perfectly with all the special cases that have been worked out in the literature. The case of triangular gaps of size two near a zig-zag boundary was seen in \cite{\sc} to be given by a formula analogous to (\ebe), in which one takes the correlation of the collection of gaps together with their mirror images, and extracts the square root from it. The case of a single triangular gap of size two near an open lattice line boundary was treated in \cite{\free}, where it was shown that the correlation has asymptotics given by the square root of the pair of gaps consisting of the original gap and its mirror image in the boundary. The case of a gap of size two near a $60^\circ$ (resp., $120^\circ$) angle with constrained zig-zag boundary was solved in \cite{\anglep} (resp, \cite{\aanglep}). Formula (\ebe) adds to this small collection of known cases a new, more complex instance, and lends support to the conjectural answer in the general case discussed in Section~8.

This is the main point we make in this paper in terms of physical interpretation, showing that the physical analogy, modeled by electrostatics in \cite{\sc}\cite{\ec}\cite{\ef}\cite{\ov}\cite{\free}\cite{\anglep}, becomes even closer, in the context of the interaction of gaps with boundaries, by considering the steady state heat flow problem in a uniform block of material.

%

\mysec{3. A quartered hexagon with dents and free boundary}


\topinsert
\centerline{\mypic{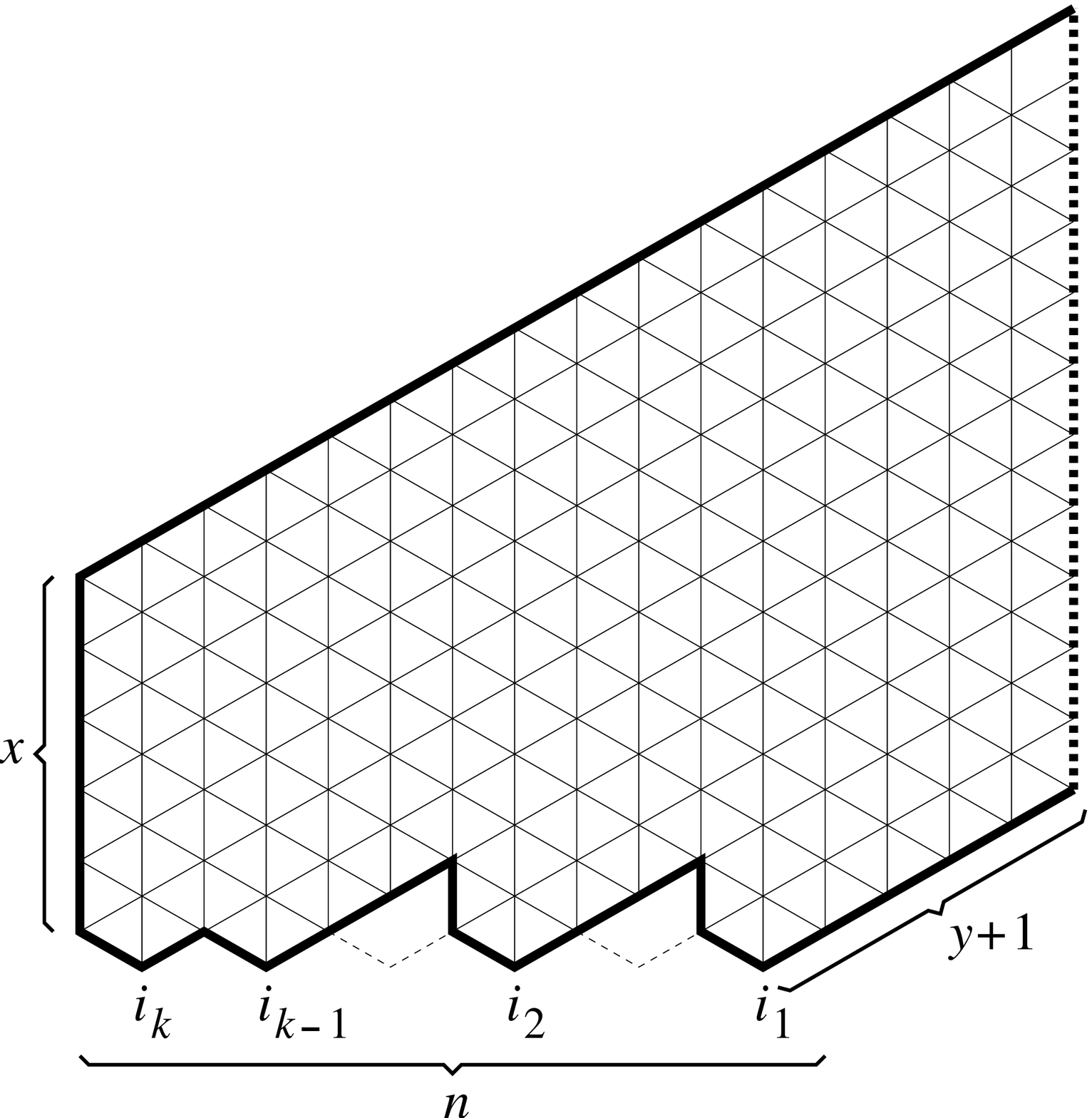}}
\medskip
\centerline{Figure~{\fcb}. {\rm $D_{n,x,y}^{i_1,\dotsc,i_k}$ for $n=6$, $x=5$, $y=4$, $k=4$, $i_1=1$, $i_2=3$, $i_3=5$, $i_4=6$.}}
\endinsert


In order to prove Theorem {\tba}, we need to enumerate the lozenge tilings of the following regions constructed from the regions $D_{n,x,y}$ defined at the beginning of Section 2. (Note that for $y=0$, this region is a quarter of the hexagon of sides $2n$, $2n$, $x$, $2n$, $2n$, $x$.) Let $0\leq k\leq n$ and $1\leq i_1<\cdots<i_k\leq n$ be integers. Then we define $D_{n,x,y}^{i_1,\dotsc,i_k}$ to be the region obtained from $D_{n,x,y}$ by making dents along the bottom zig-zag boundary as indicated in Figure {\fcb}. More precisely, label the bumps on the bottom of $D_{n,x,y}$, from right to left, by $1,\dotsc,n$, and make dents of the kind indicated in Figure {\fcb} at all bumps whose label is {\it not} in the set $\{i_1,\dotsc,i_k\}$. We denote the resulting region by $D_{n,x,y}^{i_1,\dotsc,i_k}$.

The main result of this section is the following.

\proclaim{Proposition \tca} For any integers $n,x\geq0$ and $y\geq-1$, and for any integers $1\leq i_1<\cdots<i_k\leq n$, we have
$$
\M_f(D_{n,x,y}^{i_1,\dotsc,i_k})=
\prod_{a=1}^k{x+y+n+i_a \choose y+2i_a}
\prod_{1\leq a<b\leq k}\frac{i_b-i_a}{y+i_b+i_a}.
\tag\eca
$$

\endproclaim

We point out that the analogous result from \cite{\anglep} (see Proposition 3.1 there) only covered the case of two dents, yet the corresponding product formula has a quadratic factor in it. Unlike in that case, whose proof was based on Kuo's graphical condensation method (see \cite{\Kuo}), we deduce the above result from Schur's Pfaffian identity.

We recall the definition of the Pfaffian of a skew symmetric matrix $A=\left(a_{ij}\right)_{i,j=1}^{n}$ where $n$ is even. Let ${\Cal F}_n$ be the set of perfect matchings of $(1,2,\dotsc,n)$. By convention, we always write the edges of these perfect matchings as $(i,j)$ with $i<j$. Two edges $(i,j)$ and $(k,l)$ of a given perfect matching in ${\Cal F}_n$ are said to be {\it crossed} if $i<k<j<l$ or $k<i<l<j$. Define the sign $\sgn(\mu)$ of a perfect matching $\mu$ to be $(-1)^k$, where $k$ is the number of crossed edges in $\mu$.

Then the Pfaffian of the upper triangular array $A=\left(a_{ij}\right)_{1\leq i<j\leq n}$ (or equivalently of the $n\times n$ skew symmetric matrix it determines) is defined by
$$
\Pf(A)=\sum_{\mu\in{\Cal F}_n}\sgn(\mu)\sum_{(i,j)\in\mu}a_{ij}.
\tag\ecb
$$

In the proof of Proposition {\tca} we will use the following classical identity due to Schur \cite{\Schur}. 

\proclaim{Theorem {\tcb} (Schur's Pfaffian Identity)} Let $n$ be even, and let $x_1,\dotsc,x_n$ be indeterminates. Then we have

$$
\Pf\left[\frac{x_j-x_i}{x_j+x_i}\right]_{i,j=1}^n=\prod_{1\leq i<j\leq n}\frac{x_j-x_i}{x_j+x_i}.
\tag\ecc
$$

\endproclaim

We will also use the following elementary lemma.

\proclaim{Lemma \tcc}
$(${\rm a}$)$. For any non-negative integers $k$ and $m$ we have
$$
\sum_{i=0}^m {i\choose k}={m+1\choose k+1}.
\tag\ecca
$$

$(${\rm b}$)$. For any integers $1\leq k<l\leq m$ and any non-negative integer $y$ we have
$$
\spreadlines{3\jot}
\align
&\!\!\!\!\!\!\!\!
\sum_{1\leq i<j\leq m}
{y+k+i\choose y+2k}{y+l+j\choose y+2l}-{y+k+j\choose y+2k}{y+l+i\choose y+2l}
\\
&\ \ \ \ \ \ \ \ 
=  
\frac{l-k}{y+l+k+1}\frac{m-k+1}{y+2k+1}\frac{m-l+1}{y+2l+1}
{y+m+k+1\choose y+2k}{y+m+l+1\choose y+2l}
\tag\eccab
\\
&\ \ \ \ \ \ \ \ 
=  
\frac{l-k}{y+l+k+1}
{y+m+k+1\choose m-k}{y+m+l+1\choose m-l}.
\tag\eccb
\endalign
$$

\endproclaim

\pf (a). This follows readily by repeatedly applying the basic recurrence for the binomial coefficients.

(b). Since the summand (\eccb) is zero for $i=j$, using part (a) we can write the left hand side of (\eccb) as
$$
\spreadlines{3\jot}
\align
&
\sum_{1\leq i\leq j\leq m}
{y+k+i\choose y+2k}{y+l+j\choose y+2l}-{y+k+j\choose y+2k}{y+l+i\choose y+2l}
\\
&\ \ \ \ \ \ \ \ 
=\sum_{1\leq j\leq m}
{y+k+j+1\choose y+2k+1}{y+l+j\choose y+2l}-{y+l+j+1\choose y+2l+1}{y+k+j\choose y+2k}
\\
&
=
\frac{l-k}{(y+2k+1)(y+2l+1)}
\sum_{1\leq j\leq m}
(y+2j+1){y+k+j\choose y+2k}{y+l+j\choose y+2l}.
\tag\eccc
\endalign
$$
Writing
$$
\spreadlines{3\jot}
\align
&
(y+2j+1){y+k+j\choose y+2k}{y+l+j\choose y+2l}
\\
&\ \ \ \ \ \ \ \ 
=
\frac{(j+1-l)(j+1-k)}{y+k+l+1}{y+k+j+1\choose y+2k}{y+l+j+1\choose y+2l}
\\
&\ \ \ \ \ \ \ \ \ \ \ \ 
-
\frac{(j-l)(j-k)}{y+k+l+1}{y+k+j\choose y+2k}{y+l+j\choose y+2l},
\endalign
$$
we see by telescoping cancellations that the sum on the right hand side of (\eccc) evaluates as
$$
\spreadlines{3\jot}
\align
&
\sum_{1\leq j\leq m}
(y+2j+1){y+k+j\choose y+2k}{y+l+j\choose y+2l}
\\
&\ \ \ \ \ \ \ \ 
=
\frac{(m+1-k)(m+1-l)}{y+k+l+1}{y+m+k+1\choose y+2k}{y+m+l+1\choose y+2l}.
\tag\eccd
\endalign
$$
Equation (\eccab) follows by (\eccc) and (\eccd). Equation (\eccb) follows from (\eccab) by absorbing the second and third fractions into the binomial coefficients. \epf

{\it Proof of Proposition \tca}. Each lozenge tiling of $D_{n,x,y}^{i_1,\dotsc,i_k}$ can be viewed as a family of $k$ non-intersecting paths of lozenges, as indicated in Figure {\fcc} (see also \cite{\DT}). Note that the starting points of these paths of lozenges are always the same (the $k$ northeast facing edges on the bottom, shown in thick lines in Figure {\fcd}), but the ending points can vary --- they form $k$-element subsets of the set of $n+x$ unit segments shown in thick lines on the right in Figure {\fcc}.

\topinsert
\twolinetwo{\mypic{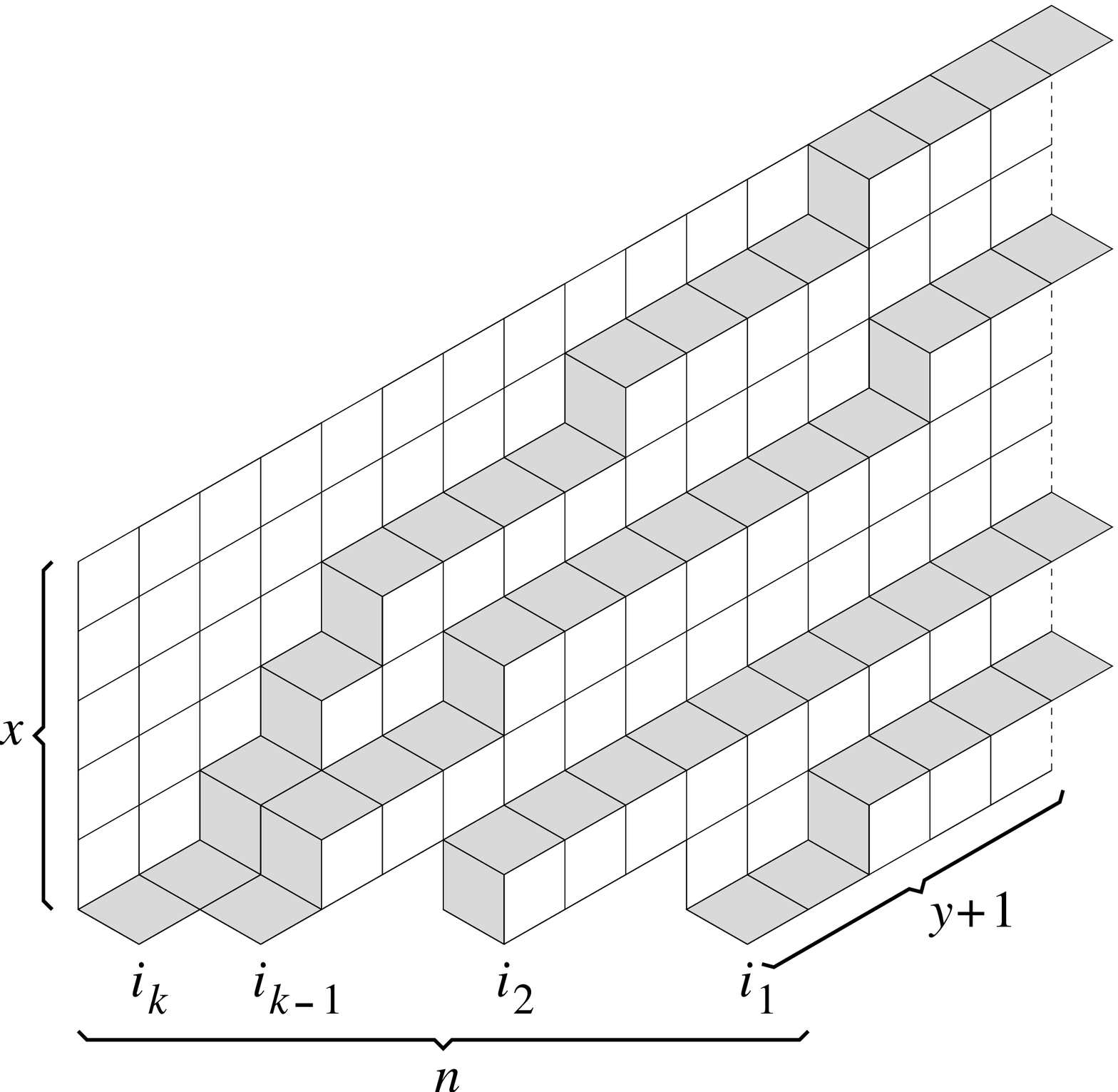}}{\mypic{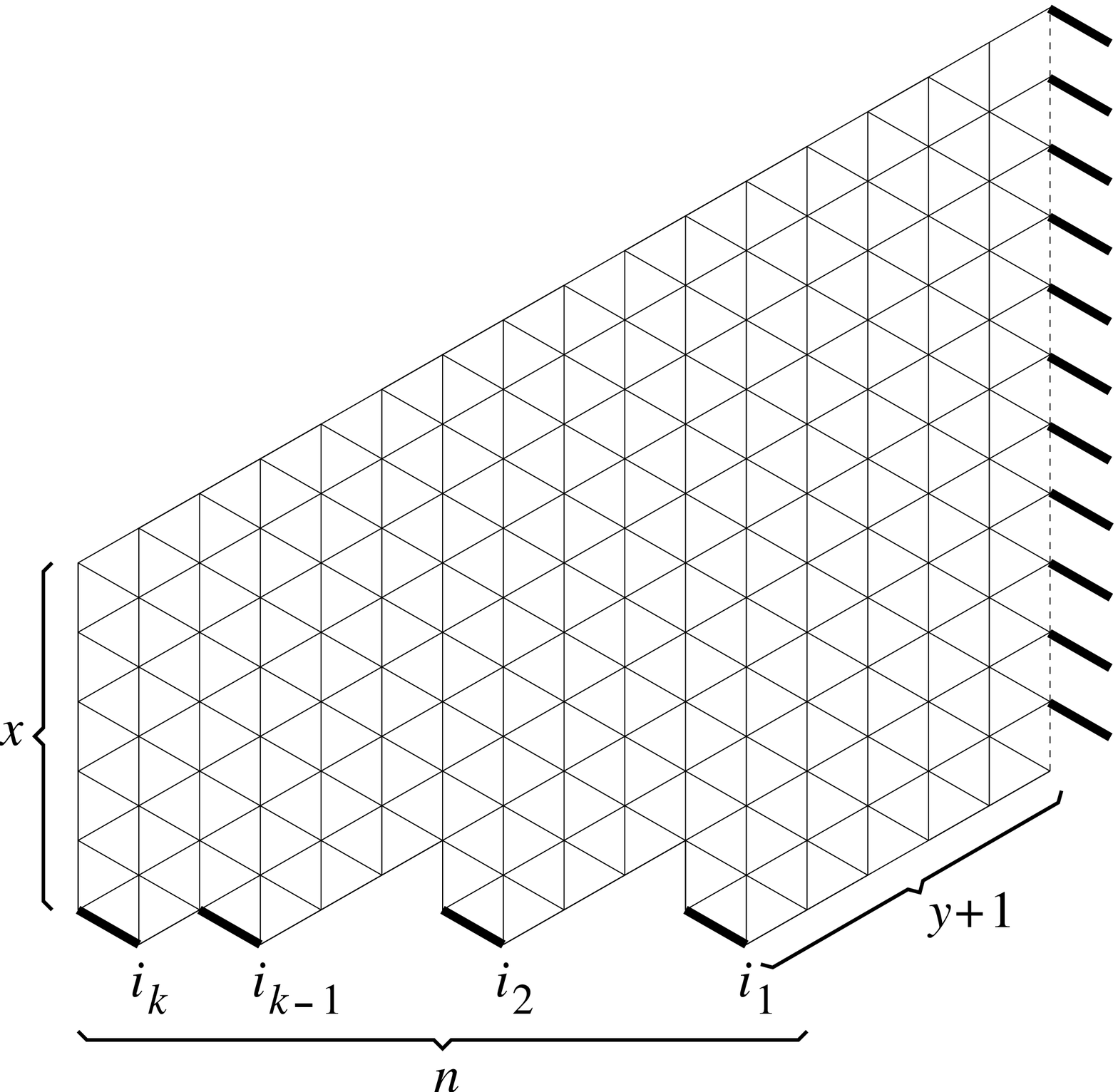}}
\medskip
\twoline{Figure~{\fcc}. {\rm Tilings and paths.}}{Figure~{\fcd}. {\rm Starting and ending segments.}}
\endinsert

These families of non-intersecting paths of lozenges can be regarded as families of non-intersecting lattice paths on $\Z^2$. Indeed, consider the affinely deformed square grid indicated in medium thick likes in Figure {\fce}. It is apparent that the shaded family of paths of lozenges is equivalent to the family of lattice paths on $\Z^2$ indicated in the figure by thick lines. These lattice paths start at the $k$ points corresponding to the marked segments along the bottom boundary in Figure {\fcd}, end at some $k$-element subset of the lattice points corresponding to the $n+x$ marked segments along the eastern boundary in Figure {\fcd}, and are allowed to take unit steps in the northern and northeastern directions.

\topinsert
\centerline{\mypic{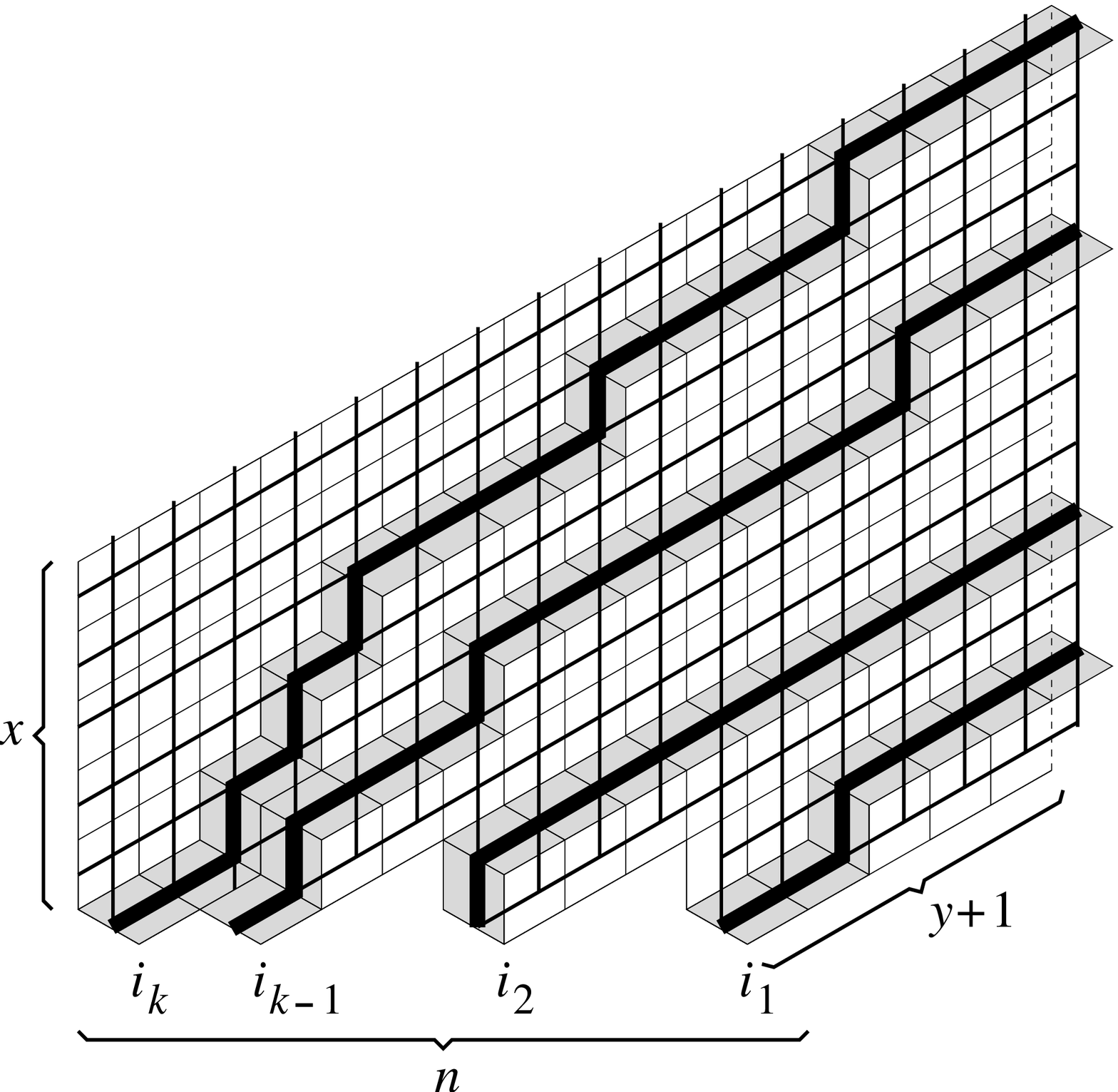}}
\medskip
\centerline{Figure~{\fce}. {\rm Regarding the paths of lozenges as lattice paths in $\Z^2$.}}
\endinsert

Coordinatize $\Z^2$ so that the origin is at the lattice point corresponding to the bottommost marked segment along the eastern boundary in Figure {\fcd}, and the $x$- and $y$-axes run in the polar directions $\pi/6$ and $\pi/2$, respectively. Then one readily sees that the starting points for the families of non-intersecting lattice paths are $A_i=(-y-2i+1,i-1)$, $i\in\{i_1,\dotsc,i_k\}$, while the possible ending points are $B_j=(0,j-1)$, $j=1,\dotsc,n+x$.
The number of families of such non-intersecting lattice paths on $\Z^2$ --- with fixed starting points, but ending points that can vary over a given set --- can be expressed as a Pfaffian using Stembridge's theorem \cite{\Ste, Theorem\,3.1}. The resulting Pfaffian has slightly different forms for different parities of $k$. 

For even $k$ we obtain that
$$
\M_f(D_{n,x,y}^{i_1,\dotsc,i_k})=\Pf\left[Q_{a,b}\right]_{1\leq a<b\leq k},
\tag\ecd
$$
where
$$
Q_{a,b}=\sum_{1\leq s<t\leq n+x}
{y+i_a+s-1\choose y+2i_a-1}{y+i_b+t-1\choose y+2i_b-1}
-
{y+i_a+t-1\choose y+2i_a-1}{y+i_b+s-1\choose y+2i_b-1}.
\tag\ece
$$

On the other hand, for $k$ odd we obtain that
$$
\M_f(D_{n,x,y}^{i_1,\dotsc,i_k})=\Pf\left[Q_{a,b}\right]_{0\leq a<b\leq k},
\tag\ecf
$$
where, for $1\leq a<b\leq k$, $Q_{a,b}$ is still given by (\ece), while 
$$
Q_{0,b}=\sum_{1\leq s\leq n+x}
{y+i_b+s-1\choose y+2i_b-1}.
\tag\ecg
$$

Quite conveniently, it turns out that both the single sum in (\ecg) and the double sum in (\ece) have simple closed form evaluations. Namely, it follows by Lemma {\tcc} that
$$
\sum_{1\leq s\leq n+x}
{y+i_b+s-1\choose y+2i_b-1}
=
{y+n+x+i_b\choose y+2i_b}
\tag\ech
$$
and
$$
\spreadlines{3\jot}
\align
&
\sum_{1\leq s<t\leq n+x}
{y+i_a+s-1\choose y+2i_a-1}{y+i_b+t-1\choose y+2i_b-1}
-
{y+i_a+t-1\choose y+2i_a-1}{y+i_b+s-1\choose y+2i_b-1}
\\
&\ \ \ \ \ \ \ \ \ \ \ \ \ \ 
=
\frac{i_b-i_a}{y+i_a+i_b}
{y+n+x+i_a\choose n+x-i_a}{y+n+x+i_b\choose n+x-i_b}.
\tag\eci
\endalign
$$

Suppose $k$ is even. By (\ecd), (\ece) and (\eci), we have that
$$
\M_f(D_{n,x,y}^{i_1,\dotsc,i_k})=
\Pf
\left[
\frac{i_b-i_a}{y+i_a+i_b}
{y+n+x+i_a\choose n+x-i_a}{y+n+x+i_b\choose n+x-i_b}
\right]_{1\leq a<b\leq k}.
\tag\ecj
$$
Using the identity
$$
\Pf[x_ix_ja_{ij}]=x_1\cdots x_n\Pf[a_{ij}]\tag\eck
$$
(see e.g. \cite{\Ste, Proposition\,2.3(a)}), we obtain from (\ecj) that
$$
\M_f(D_{n,x,y}^{i_1,\dotsc,i_k})=
\prod_{a=1}^k{y+n+x+i_a\choose n+x-i_a}
\Pf
\left[
\frac{i_b-i_a}{y+i_a+i_b}
\right]_{1\leq a<b\leq k}.
\tag\ecl
$$
Equation (\eca) follows from (\ecl) by using Schur's Pfaffian indentity (\ecc), thus proving the proposition in the case of even $k$.

The case when $k$ is odd follows similarly, by using (\ecf), (\ece), (\ecg), (\ech), (\eci) and the  $x_1=0$ specialization of Schur's Pfaffian indentity (\ecc). \epf


\topinsert
\centerline{\mypic{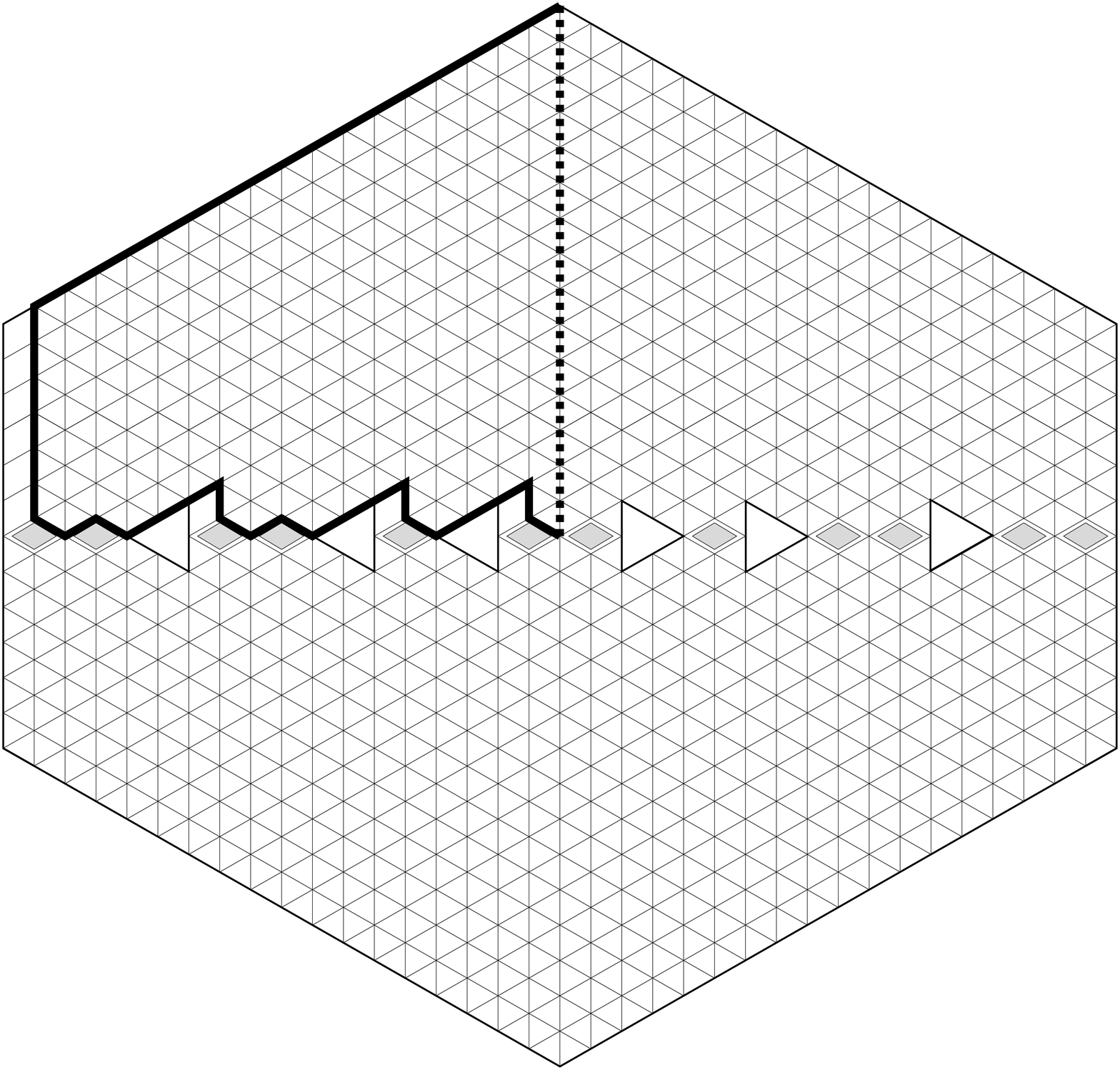}}
\medskip
\centerline{Figure~{\fcf}. {\rm Generalization of SSC plane partitions, even by even by even case.}}
\endinsert

\topinsert
\centerline{\mypic{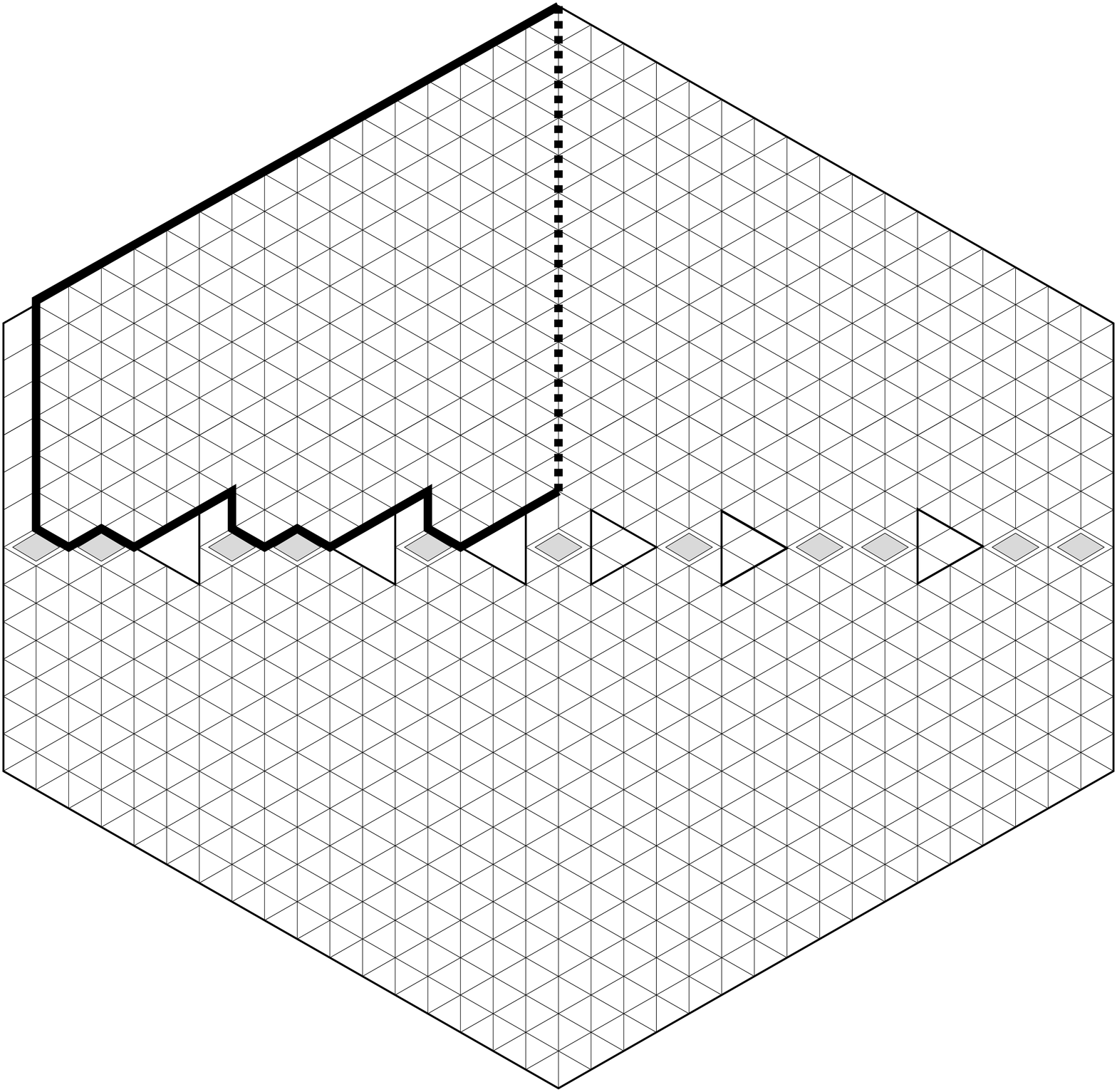}}
\medskip
\centerline{Figure~{\fcg}. {\rm Generalization of SSC plane partitions, even by odd by odd case.}}
\endinsert

Consider the hexagon $H_{2n,2n,2x}$ of sides $2n$, $2n$, $2x$, $2n$, $2n$, $2x$ (clockwise from the northwestern side) on the triangular lattice, and let $\ell_h$ and $\ell_v$ be its horizontal and vertical symmetry axes. Note that there are $n$ different positions one can place a left-pointing lattice triangle of side two symmetrically about $\ell_h$, to the left of $\ell_v$. Label them from right to left by $1,\dotsc,n$, and call them {\it slots}. Denote by $H_{2n,2n,2x}(k_1,\dotsc,k_s)$ the region obtained from $H_{2n,2n,2x}$ by removing from it left-pointing triangles of side two from slots $k_1,\dotsc,k_s$, as well as their mirror images in $\ell_v$ (see Figure {\fcf}). Let $H_{2n+1,2n+1,2x}(k_1,\dotsc,k_s)$ be the region obtained analogously from the hexagon $H_{2n+1,2n+1,2x}$ of sides $2n+1$, $2n+1$, $2x$, $2n+1$, $2n+1$, $2x$ (clockwise from the northwestern side; see Figure {\fcg}).

For a lattice region $R$ that has both a horizontal and a vertical symmetry axis, denote by $\M_{-,|}(R)$ the number of those lozenge tilings of $R$ that are invariant under reflection across both the horizontal and vertical symmetry axis of $R$.

Then Theorem {\tca} implies the following result (to which it is in fact equivalent).

\proclaim{Corollary {\tcd} (Generalization of SSC plane partitions)} $(${\rm a}$)$. Let $n,x\geq0$ and $1\leq k_1<\cdots<k_s\leq n$ be integers. If $k_1>1$ set $t=0$, otherwise define $t$ by requiring $k_i-i=0$, $i=1,\dotsc,t$, and $k_{t+1}-(t+1)>0$. Let $\{1,\dotsc,n\}\setminus\{k_1,\dotsc,k_s\}=\{i_1,\dotsc,i_r\}$.

Then we have:

$(${\rm a}$)$.
$$
\spreadlines{3\jot}
\align
\M_{-,|}(H_{2n,2n,2x}(k_1,\dotsc,k_s))
&=
\M_f(D_{n,x,2t-1}^{i_1,\dotsc,i_r})
\\
&=
\prod_{a=1}^r{x+2t+n+i_a-1 \choose 2t+2i_a-1}
\prod_{1\leq a<b\leq r}\frac{i_b-i_a}{2t+i_a+i_b-1}.
\\
\tag\ecm
\endalign
$$

$(${\rm b}$)$.
$$
\spreadlines{3\jot}
\align
\M_{-,|}(H_{2n+1,2n+1,2x}(k_1,\dotsc,k_s))
&=
\M_f(D_{n,x,2t}^{i_1,\dotsc,i_r})
\\
&=
\prod_{a=1}^r{x+2t+n+i_a \choose 2t+2i_a}
\prod_{1\leq a<b\leq r}\frac{i_b-i_a}{2t+i_a+i_b}.
\\
\tag\ecn
\endalign
$$

\endproclaim

\pf For part (a), note that any lozenge tiling of $H_{2n,2n,2x}(k_1,\dotsc,k_s)$ which is symmetric across the horizontal must contain the $2n-2s$ lozenges indicated by a shading in Figure {\fcf}. When removing these lozenges, $H_{2n,2n,2x}(k_1,\dotsc,k_s)$ gets disconnected into two congruent regions, one above and one below $\ell_h$; the horizontally symmetric tilings of $H_{2n,2n,2x}(k_1,\dotsc,k_s)$ are in bijection with the tilings of say the upper region.

Then the horizontally {\it and vertically} symmetric tilings of $H_{2n,2n,2x}(k_1,\dotsc,k_s)$ are in bijection with the tilings of the upper region which are symmetric across $\ell_v$. In turn, after removing the forced lozenges from the upper region, these tilings are readily seen to be in bijection with the lozenge tilings of the left half of the upper region, provided its boundary along $\ell_v$ is considered free. However, this is precisely the region on the right hand side of (\ecm). The formula follows then by Theorem {\tca}.

Equation (\ecn) is proved by precisely the same argument. \epf

\medskip
\flushpar
{\smc Remark 2.} The special case $s=0$ of Corollary {\tcd} corresponds to the class of symmetric and self-complementary plane partitions, which was first proved by Proctor \cite{\Proc}.

\medskip
\flushpar
{\smc Remark 3.} There are four symmetry classes for the lozenge tilings of $H_{n,n,2x}(k_1,\dotsc,k_s)$: No symmetry requirement, horizontally symmetric, vertically symmetric, and both  horizontally and vertically symmetric. Corollary {\tcd} shows that the fourth symmetry class is given by a simple product formula. Even though the other three turn out not to be given by simple product formulas, it is true that the first is equal to the product of the second and third. Namely, we have that
$$
\M(H_{n,n,2x}(k_1,\dotsc,k_s))
=
\M_{-}(H_{n,n,2x}(k_1,\dotsc,k_s))\M_{|}(H_{n,n,2x}(k_1,\dotsc,k_s)),
\tag\eco
$$
where $\M_{-}(R)$ and $\M_{|}(R)$ denote the number of horizontally, respectively vertically symmetric lozenge tilings of $R$ (see \cite{\fakt}).

\mysec{4. A limit formula for regions with two dents}





In our proof of Theorem {\tba} we will use the following result, which gives the limit of the ratio between the number of lozenge tilings of the regions $D_{n,n,0}^{[n]\setminus\{i,j\}}$ and $D_{n,n,0}^{[n]\setminus\{1,2\}}$ when $i$ and $j$ are fixed, and $n$ tends to infinity\footnote{ Recall that $[n]$ denotes the set $\{1,2,\dotsc,n\}$.}.

\proclaim{Proposition \tda} For any fixed integers $1\leq i<j$, we have
$$
\lim_{n\to\infty}
\frac{\M_f\left(D_{n,n,0}^{[n]\setminus\{i,j\}}\right)}{\M_f\left(D_{n,n,0}^{[n]\setminus\{1,2\}}\right)}
=
4\,\frac{j-i}{j+i}\frac{1}{2^{2i-2}}{2i-1\choose i-1}\frac{1}{2^{2j-2}}{2j-1\choose j-1}.
\tag\eda
$$

\endproclaim

\pf Proposition {\tca} provides an explicit formula for the numerator on the left hand side above. Write it in the form
$$
\spreadlines{3\jot}
\align
&
\M_f\left(D_{n,n,0}^{[n]\setminus\{i,j\}}\right)
=
\frac
{
{\displaystyle \prod_{a=1}^n {2n+a\choose a}}
}
{
{\displaystyle {2n+i\choose i}{2n+j\choose j}}
}
\frac{j-i}{j+i}
\\
&
\times
\frac
{\displaystyle \prod_{1\leq a<b\leq n}\dfrac{b-a}{b+a}}
{
\dfrac{(i-1)!}{(i+1)\cdots(2i-1)}
\dfrac{(n-i)!}{(2i+1)\cdots(i+n)}
\dfrac{(j-1)!}{(j+1)\cdots(2j-1)}
\dfrac{(n-j)!}{(2j+1)\cdots(j+n)}
}
\\
&
=\frac{\dfrac{j-i}{j+i}}{\displaystyle {2n+i\choose i}{2n+j\choose j}}
{\displaystyle \prod_{a=1}^n {2n+a\choose a}}
{\displaystyle \prod_{1\leq a<b\leq n}\dfrac{b-a}{b+a}}
\\
&\ \ \ \ \ \ \ \ \ \ \ \ \ \ \ \ \ \ \ \ \ \ \ \ \ \ \ \ \ \ \ \
\times
{\displaystyle {2i-1\choose i-1}{2j-1\choose j-1}}
\dfrac{(n+i)!}{(n-i)!(2i)!}\dfrac{(n+j)!}{(n-j)!(2j)!}.
\tag\edb
\endalign
$$
Dividing (\edb) by its $i=1$, $j=2$ specialization, we obtain
$$
\frac{\M_f\left(D_{n,n,0}^{[n]\setminus\{i,j\}}\right)}{\M_f\left(D_{n,n,0}^{[n]\setminus\{1,2\}}\right)}
=
\frac{\dfrac{j-i}{j+i}}{\dfrac{2-1}{2+1}}
\frac
{\displaystyle {2i-1\choose i-1}{2j-1\choose j-1}}
{\displaystyle {2\cdot 1-1\choose 1-1}{2\cdot 2-1\choose 2-1}}
\frac
{\dfrac{(n+i)!}{(n-i)!}}
{\dfrac{(n+1)!}{(n-1)!}}
\frac
{\dfrac{(2n+1)!}{(2n-1)!}}
{\dfrac{(2n+i)!}{(2n-i)!}}
\frac
{\dfrac{(n+j)!}{(n-j)!}}
{\dfrac{(n+2)!}{(n-2)!}}
\frac
{\dfrac{(2n+2)!}{(2n-2)!}}
{\dfrac{(2n+j)!}{(2n-j)!}}.
\tag\edc
$$
However, writing both the numerator and denominator as polynomials of degree $i$, one obtains that
$$
\lim_{n\to\infty}
\frac
{\dfrac{(n+i)!}{(n-i)!}}
{\dfrac{(2n+i)!}{(2n-i)!}}
=
\frac{1}{2^{2i}}.
\tag\edd
$$
Taking the $n\to\infty$ limit in (\edc) and using (\edd) four times, one obtains (\eda). \epf

\flushpar
{\smc Remark 4.} The same approach shows that, more generally, we have
$$
\lim_{n\to\infty}
\frac{\M_f\left(D_{n,n,0}^{[n]\setminus\{i_1,\dotsc,i_k\}}\right)}{\M_f\left(D_{n,n,0}^{[n]\setminus\{1,\dotsc,k\}}\right)}
=
\prod_{a=1}^n
\frac
{\dfrac{1}{2^{2i_a}}{\displaystyle {2i_a-1\choose i_a-1}}}
{\dfrac{1}{2^{2a}}{\displaystyle {2a-1\choose a-1}}}
\prod_{1\leq a<b\leq n}
\frac
{\dfrac{i_b-i_a}{i_b+i_a}}
{\dfrac{b-a}{b+a}}.
\tag\ede
$$

\mysec{5. A double sum expression for $\omega_c(\al,\be)$}


\topinsert
\centerline{\mypic{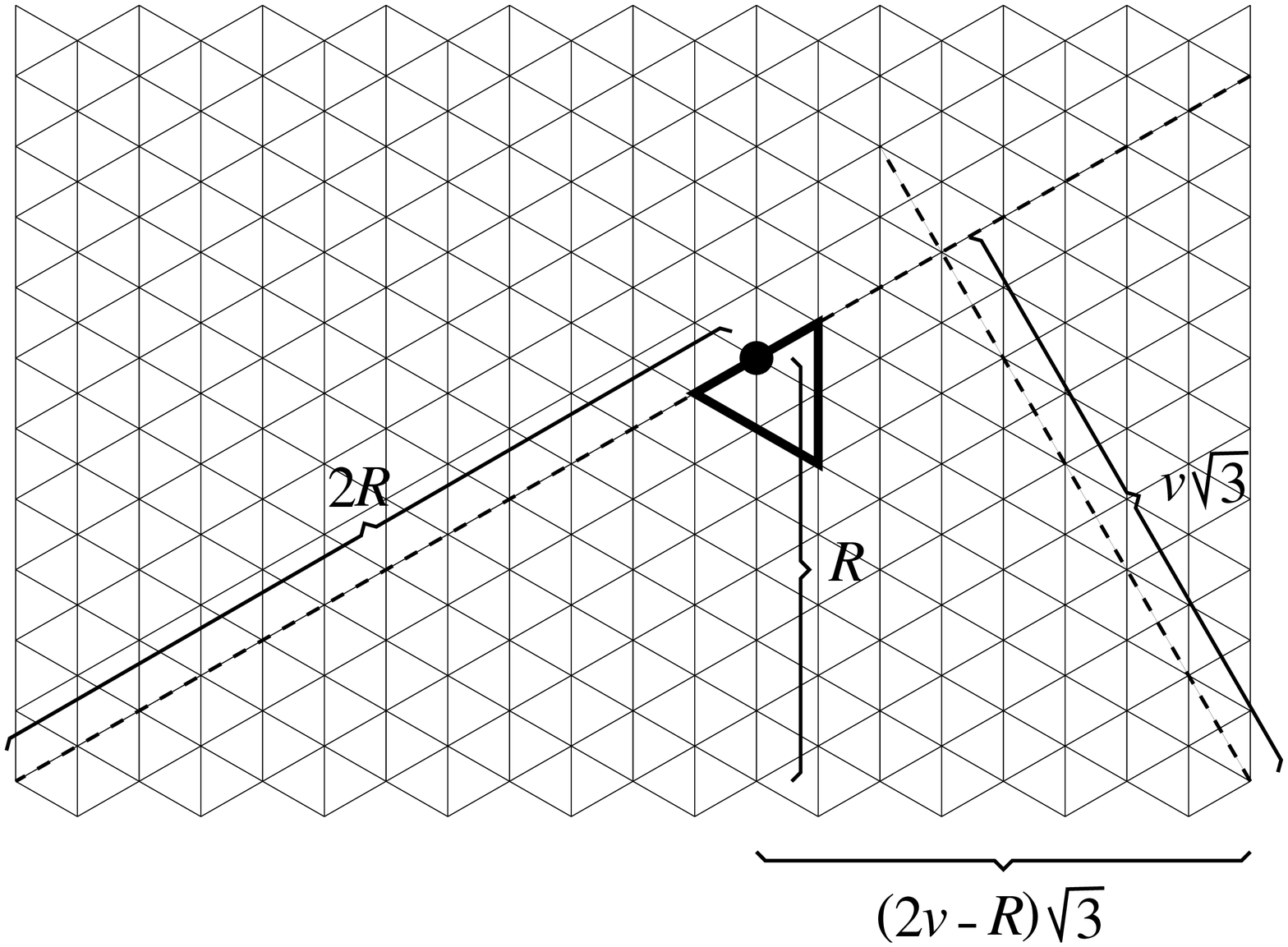}}
\medskip
\centerline{Figure~{\fea}. {\rm Changing to $(R,v)$-coordinates.}}
\endinsert

We will find it convenient to make the change of variables $\al=2v-R$, $\be=R$, where $R$ and $v$ are non-negative integers having the significance shown in Figure {\fea}. This will allow us to relate very closely one part of the calculations we need to the calculations already worked out in \cite{\anglep}, where precisely these $(R,v)$ coordinates were used.

The following double sum expression for the correlation $\omega_c(\al,\be)$ will be the starting point for our proof of Theorem {\tba}.

\proclaim{Lemma \tea} Write $\al=2v-R$, $\be=R$, with $R$ and $v$ non-negative integers. Then we have
$$
\spreadlines{4\jot}
\align
&\!\!\!\!\!\!\!\!\!\!\!\!\!\!\!\!\!\!\!
\omega_c(\al,\be)=\omega_c(2v-R,R)
\\
&\!\!\!\!\!\!\!\!\!\!\!\!\!\!\!\!\!\!
=
4R
\left|
\sum_{a=0}^R\sum_{b=0}^R
(-1)^{a+b}
\frac{(R+a-1)!\,(R+b-1)!}{(2a)!\,(R-a)!\,(2b)!\,(R-b)!}
\right.
\\
&
\times
\frac{(2v'+2a+1)!\,(2v'+2b+1)!}{2^{2(2v'+a+b)}(v'+a)!\,(v'+a+1)!\,(v'+b)!\,(v'+b+1)!}
\left.
\frac{(b-a)^2}{2v'+a+b+2}\right|,
\tag\eea
\endalign
$$
where $v'=2v-R-1$.

\endproclaim

\topinsert
\twoline{\mypic{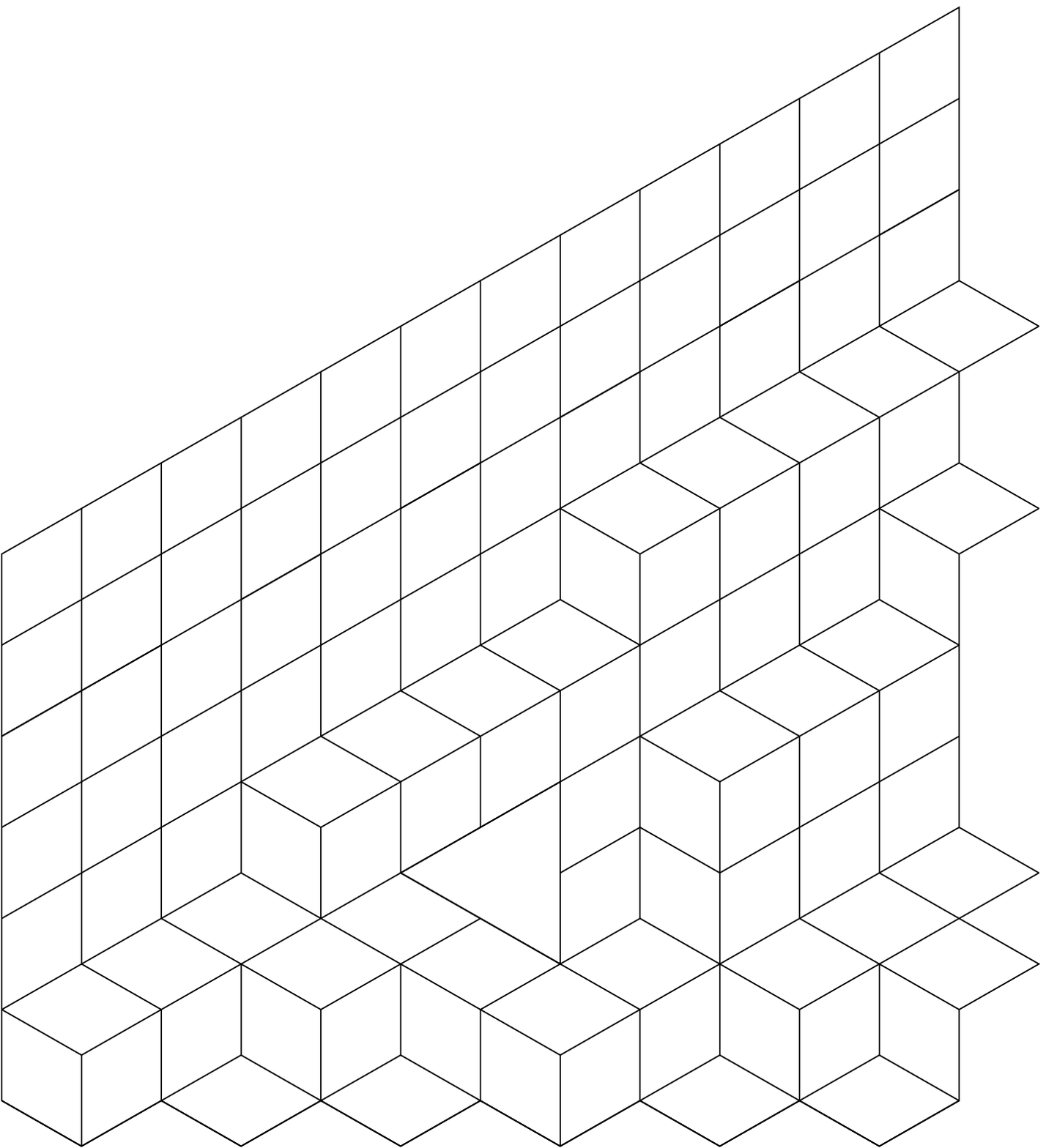}}{\mypic{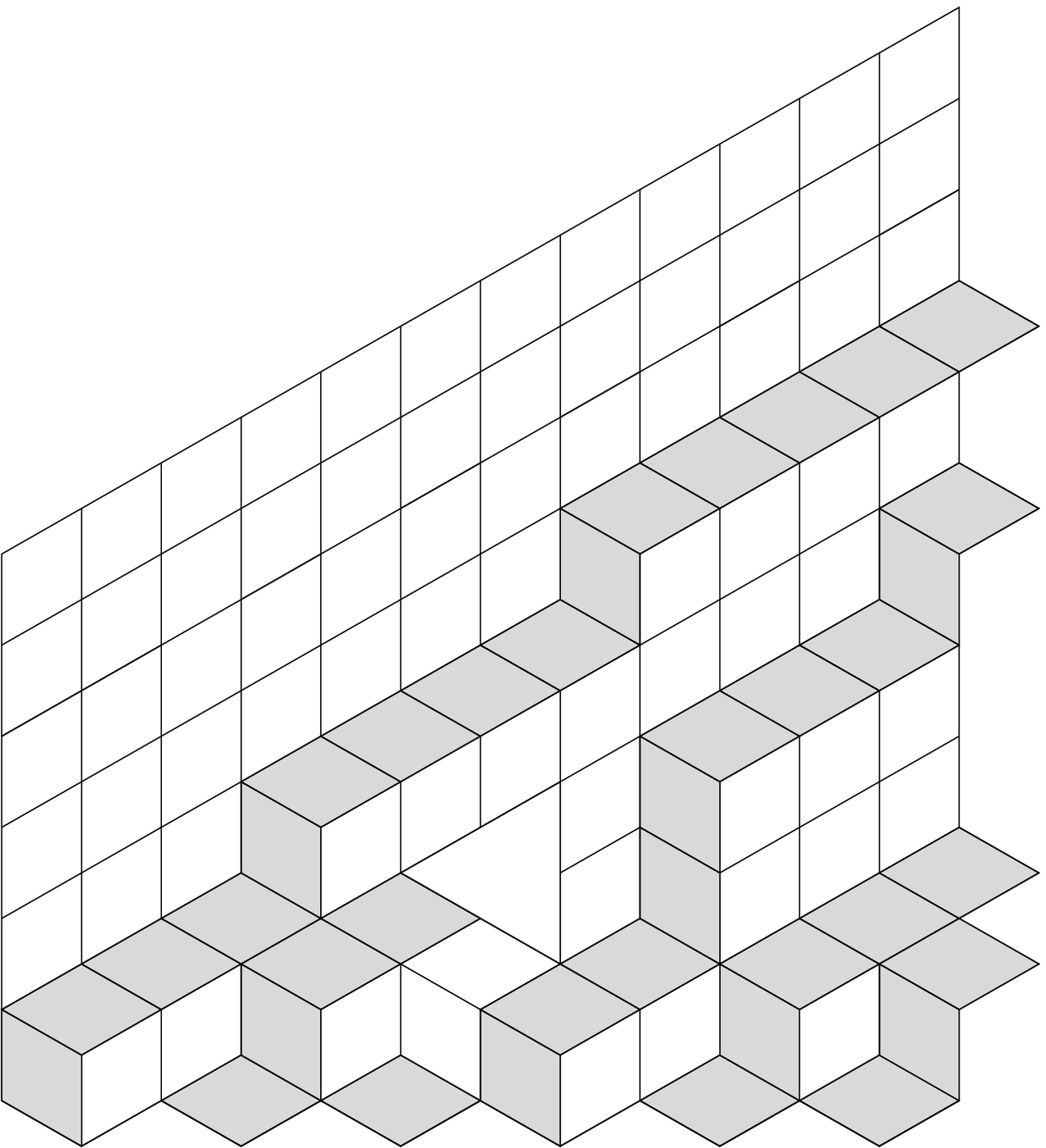}}
\medskip
\twoline{Figure~{\feb}. {\rm $D_{6,6,0}(3,3;\{1,2,6,8\})$}. }
{Figure~{\fec}. {\rm Paths of lozenges.}}
\medskip
\twoline{\mypic{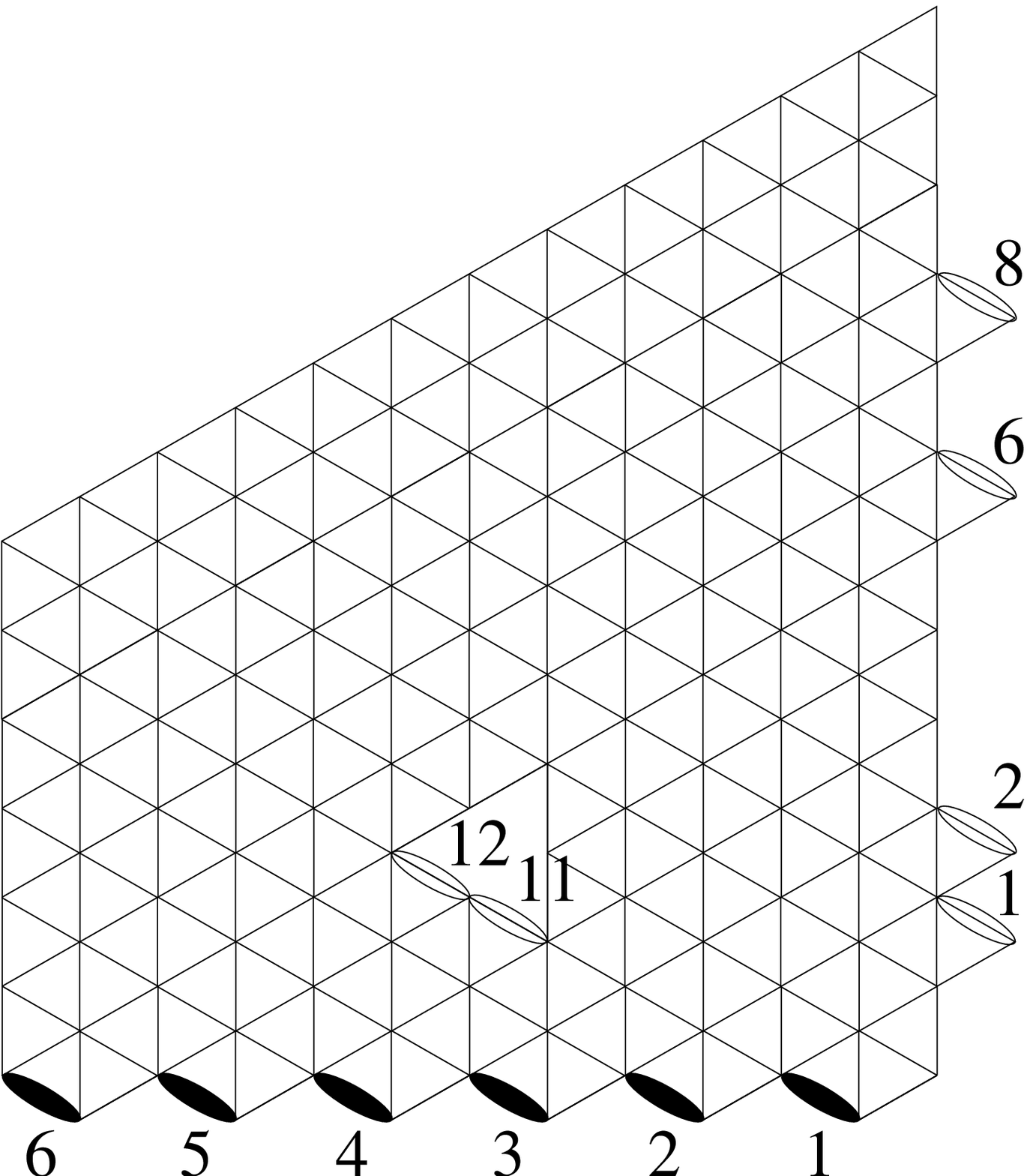}}{\mypic{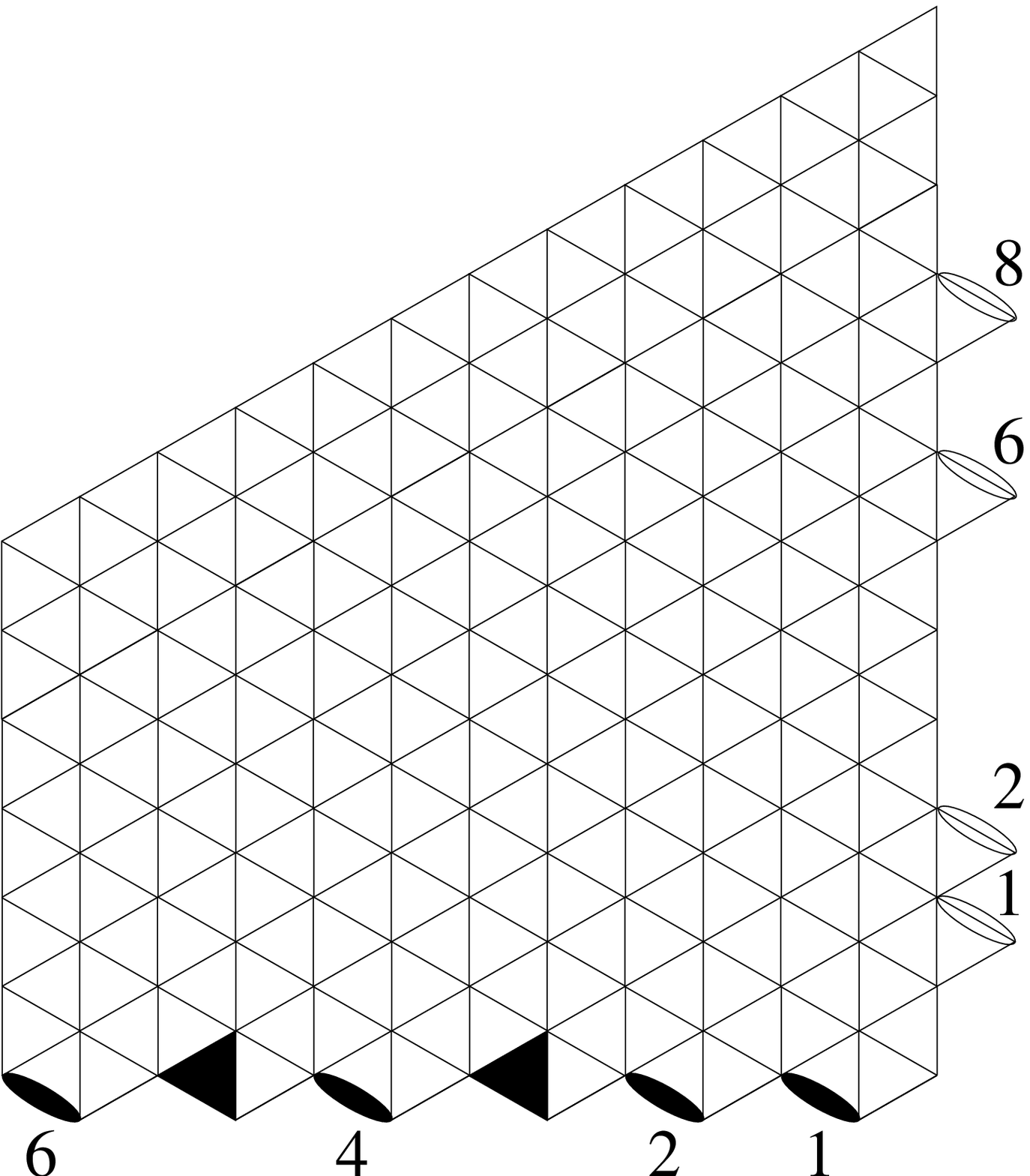}}
\medskip
\twoline{Figure~{\fed}. {\rm Labeling starting and ending points.}}
{Figure~{\fee}. {\rm $D_{6,6,0}^{1,2,4,6}(\{1,2,6,8\})$.}}
\endinsert

\pf Let $T$ be the set of $2n-2$ tile positions that straddle the free boundary on the right of the regions $D_{n,n,0}(\al,\be)$ and $D_{n,n,0}^{i_1,\dotsc,i_k}$; label them from 1 to $2n$ starting from the bottom. For any subset of $S\subset T$ with $n-2$ elements, let $D_{n,n,0}(\al,\be;S)$ and $D_{n,n,0}^{i_1,\dotsc,i_k}(S)$ be the regions obtained from $D_{n,n,0}(\al,\be)$ and $D_{n,n,0}^{i_1,\dotsc,i_k}$, respectively, by replacing the free boundary on their right sides by the lattice path which leaves all lozenges in $S$ to the left of it, and bisects all lozenges in $T\setminus S$  (see Figures {\fed} and {\fee} for two examples); note that these two new families of regions do not have free boundary anymore.

Then we have
$$
\spreadlines{3\jot}
\align
\M_f(D_{n,n,0}(\al,\be))&=\M_f(D_{n,n,0}(2v-R,R))
\\
&
=
\sum_{S\subset T\atop|S|=n-2}\M(D_{n,n,0}(2v-R,R;S)),
\tag\eeb
\endalign
$$
where, as usual, for a lattice region $R$ with constrained boundary on the triangular lattice, $\M(R)$ denotes the number of lozenge tilings of $R$.

Recall the bijection between lozenge tilings and families of non-intersecting lattice paths mentioned in Section 3 (see also Figure {\fec}). Label the starting segments of these paths from right to left by $1,\dotsc, n$. Label the possible ending segments along the free boundary on the right of $D_{n,n,0}(\al,\be)$ from bottom to top by $1,\dotsc,2n-2$, and the two ending segments along the gap by $2n-1$ and $2n$.  

Let $A$ be the $n\times 2n$ matrix
$$
A=(a_{i,j})_{1\leq i\leq n,1\leq j\leq 2n},
$$
where $a_{i,j}$ is equal to the number of paths of lozenges from starting segment $i$ to ending segment $j$. Then by the Gessel-Viennot-Lindstr\"om theorem (see e.g. \cite{\Lind}, \cite{\GV} or \cite{\Ste}) we obtain that\footnote{ We write $A_I^J$ for the minor of matrix $A$ corresponding to rows $I$ and columns $J$.}
$$
\M(D_{n,n,0}(2v-R,R;S)
=
\left|\det A_{[n]}^{S\cup\{2n-1,2n\}}\right|.
\tag\eec
$$

Doing Laplace expansion along columns $2n-1$ and $2n$ in $\det A_{[n]}^{S\cup\{2n-1,2n\}}$, we obtain that
$$
\det A_{[n]}^{S\cup\{2n-1,2n\}} = \sum_{1\leq a<b\leq n}(-1)^{a+b}
\det A_{\{a,b\}}^{\{2n-1,2n\}}
\det A_{[n]\setminus\{a,b\}}^{S}.
\tag\eed
$$

It is easy to see that the paths of rhombi ending at the ending points labeled $2n-1$ and $2n$ (i.e., ending at the gap) can only start at starting points with labels in the range from $2v-R$ to $2v$. Therefore, for $a$ or $b$ outside this range, the matrix $A_{\{a,b\}}^{\{2n-1,2n\}}$ has a zero row, and hence its determinant is zero. Thus we can rewrite (\eed) as
$$
\det A_{[n]}^{S\cup\{2n-1,2n\}} = \sum_{0\leq a<b\leq R}(-1)^{a+b}
\det A_{\{2v-R+a,2v-R+b\}}^{\{2n-1,2n\}}
\det A_{[n]\setminus\{2v-R+a,2v-R+b\}}^{S}.
\tag\eee
$$

It follows from the Gessel-Viennot-Lindstr\"om theorem that 
$$
\epsilon\det A_{[n]\setminus\{a,b\}}^{S}
=
\M(D_{n,n,0}^{[n]\setminus\{a,b\}}(S)),
\tag\eef
$$
for some sign $\epsilon=\pm1$ (see the bottom right picture in Figure {\fea}); furthermore, it is easy to see that $\epsilon$ is the same for all subsets $S$.

The entries of $A$ are easily found, being given by binomial coefficients. In particular, one has
$$
\spreadlines{4\jot}
\spreadmatrixlines{4\jot}
\align
\det A_{\{2v-R+a,2v-R+b\}}^{\{2n+1,2n+2\}}
&=
\det
\left[\matrix
{R+a-1\choose 2a} & {R+b-1\choose 2b}\\
{R+a-1\choose 2a-1} & {R+b-1\choose 2b-1}
\endmatrix\right]
\\
&=
\frac{2R(b-a)(R+a-1)!\,(R+b-1)!}{(2a)!\,(R-a)!\,(2b)!\,(R-b)!}.
\tag\eeg
\endalign
$$

By equations (\eeb)-(\eeg) we obtain that
$$
\spreadlines{3\jot}
\align
&
\M_f(D_{n,n,0}(\al,\be))=\M_f(D_{n,n,0}(2v-R,R))
\\
&\!\!\!\!\!\!\!\!
=
2R
\left|
\sum_{S\subset T\atop|S|=n-2}\sum_{0\leq a<b\leq R}
(-1)^{a+b}\frac{(b-a)(R+a-1)!\,(R+b-1)!}{(2a)!\,(R-a)!\,(2b)!\,(R-b)!}
\M(D_{n,n,0}^{[n]\setminus\{2v-R+a,2v-R+b\}}(S))
\right|
\\
&\!\!\!\!\!\!\!\!
=
2R
\left|
\sum_{0\leq a<b\leq R}
(-1)^{a+b}\frac{(b-a)(R+a-1)!\,(R+b-1)!}{(2a)!\,(R-a)!\,(2b)!\,(R-b)!}
\sum_{S\subset T\atop|S|=n-2}
\M(D_{n,n,0}^{[n]\setminus\{2v-R+a,2v-R+b\}}(S))
\right|
\\
&\!\!\!\!\!\!\!\!
=
2R
\left|
\sum_{0\leq a<b\leq R}
(-1)^{a+b}\frac{(b-a)(R+a-1)!\,(R+b-1)!}{(2a)!\,(R-a)!\,(2b)!\,(R-b)!}
\M_f(D_{n,n,0}^{[n]\setminus\{2v-R+a,2v-R+b\}})
\right|.
\tag\eeh
\endalign
$$

Divide both sides of (\eeh) by $\M_f(D_{n,n,0}^{[n]\setminus\{1,2\}})$, and take the limit as $n\to\infty$. Then by Lemma {\tda} we obtain that
$$
\spreadlines{3\jot}
\align
&
\omega_c(\al,\be)=\omega_c(2v-R,R)=
\lim_{n\to\infty}\frac{\M_f(D_{n,n,0}(2v-R,R))}{\M_f(D_{n,n,0}^{[n]\setminus\{1,2\}})}\\
&\!\!\!\!\!\!\!\!\!\!\!\!\!\!\!\!\!\!\!\!
=
8R
\left|
\sum_{0\leq a<b\leq R}
(-1)^{a+b}\frac{(R+a-1)!\,(R+b-1)!}{(2a)!\,(R-a)!\,(2b)!\,(R-b)!}
\right.
\\
&
\left.
\times
\frac{(2v'+2a+1)!}{2^{2(v'+a)}(v'+a)!(v'+a+1)!}
\frac{(2v'+2b+1)!}{2^{2(v'+b)}(v'+b)!(v'+b+1)!}
\frac{(b-a)^2}{2v'+a+b+2}\right|,
\tag\eei
\endalign
$$
where $v'=2v-R-1$.

Note that the summand above becomes zero when $a=b$, and is invariant under swapping $a$ and $b$. This leads to (\eea). \epf


\mysec{6. Reduction of the double sum to simple sums}

As in \cite{\sc,\S6}, define the moment sums $T^{(k)}(R,v;x)$, for non-negative integers $k$ and real $x\in[0,1]$, by
$$
T^{(k)}(R,v;x):=
\frac{1}{R}
\sum_{a=0}^R\frac{(-R)_a(R)_a(3/2)_{v+a}}{(1)_a(1/2)_a(2)_{v+a}}\left(\frac{x}{4}\right)^aa^k.
\tag\efa
$$

The double sum expression for $\omega_c$ given in the previous section can be written in terms of these moments (which are simple sums) as follows.

\proclaim{Lemma \tfa} We have that
$$
\spreadlines{3\jot}
\align
&
\omega_c(2v-R,R)=
\\
&\ \ \ \ 
8R
\left|
\int_0^1 T^{(2)}(R,v';x)T^{(0)}(R,v';x)x^{2v'+1} dx
-
\int_0^1 \left(T^{(1)}(R,v';x)\right)^2x^{2v'+1} dx
\right|,
\tag\efb
\endalign
$$
where $v'=2v-R-1$.

\endproclaim

\pf Were it not for the denominator $2v'+a+b+2$ in the summand of (\eea), the double sum
(\eea) would separate and could be expressed in terms of simple sums. A good way to overcome this obstacle is to write
$$
\frac{1}{2v'+a+b+2}=\int_0^1 x^{2v'+a+b+1}\,dx
\tag\efc
$$
(a trick we used also in \cite{\ri} and \cite{\sc}). Indeed, interchanging then the order of summation and integration, the double sum becomes an integral of a double sum that separates.

The details of the calculation are as follows. One readily verifies that we have
$$
\frac{(R+a-1)!}{(2a)!\,(R-a)!}
=
\frac{(-1)^a (-R)_a(R)_a}{4^a R\, (1)_a(1/2)_a}
\tag\efd
$$
and
$$
\frac{(2v+2a+1)!}{2^{2v+2a}(v+a)!(v+a+1)!}
=
\frac{(3/2)_{v+a}}{(2)_{v+a}}.
\tag\efe
$$

Expand out the factor $(b-a)^2$ in the summand in (\eea), writing it as $a^2-2ab+b^2$.
Using (\efa) and (\efc)-(\efe), we see that the double sum in (\eea) separates, and can be written as
$$
\spreadlines{3\jot}
\align
\int_0^1 T^{(2)}(R,v';x)T^{(0)}(R,v';x)x^{2v'+1} dx
&-
2
\int_0^1 T^{(1)}(R,v';x)T^{(1)}(R,v';x)x^{2v'+1} dx
\\
&+
\int_0^1 T^{(0)}(R,v';x)T^{(2)}(R,v';x)x^{2v'+1} dx,
\endalign
$$
which together with (\eea) yields the expression given in (\efb). \epf

\mysec{7. The asymptotics of the integrals in Lemma {\tfa}. Proof of Theorem {\tba}.}

Recall that we need the asymptotics of $\omega_c(\al,\be)$ as $\al$ and $\be$ approach infinity so that $\al=q\be$, where $q>0$ is a fixed rational number. Since in terms of the coordinates $(R,v)$ we have $\al=2v-R$ and $\be=R$ (see the beginning of the previous section), the relationship between $R$ and $v'=2v-R-1$ is
$$
2v-R=qR\Longleftrightarrow v'=qR-1.\tag\ega
$$
By Proposition 7.1 of \cite{\sc}, for any integer $n\geq0$ we have
$$
\spreadlines{3\jot}
\align
&\!\!\!\!\!\!\!\!\!
\left|T^{(n)}(R,qR-1;x)-\frac{2}{\sqrt{\pi}}\frac{1}{\root 4\of {q^2+\frac{x}{4-x}}}
\frac{1}{R^{3/2}}\left(R\sqrt{\frac{x}{4-x}}\,\right)^n
\right.
\\
&
\left.
\phantom{\frac{1}{\root 4\of {q^2+\frac{x}{4-x}}}}
\times
\cos\left[R\arccos\left(1-\frac x2\right)-\frac12\arctan\frac1q\sqrt{\frac{x}{4-x}}+\frac{n\pi}{2}
\right]\right|\leq MR^{n-5/2}\tag\egb
\endalign
$$
for $R\geq R_0$, where $R_0$ and $M$ are independent of $x\in(0,1]$.

As shown in Section 8 of \cite{\sc}, the asymptotics of the integrals in Lemma {\tfa}, as $v'=qR-1$ and $R\to\infty$, is unchanged if the moments $T^{(n)}(R,qR-1;x)$ are replaced by their approximants from equation (\egb). Using also the formula $\cos u\cos v=(\cos(u+v)+\cos(u-v))/2$ for the product of two cosines that arises, we obtain therefore that
$$
\spreadlines{3\jot}
\align
&\!\!\!\!
\int_0^1 T^{(2)}(R,v';x)T^{(0)}(R,v';x)x^{2v'+1} dx
\\
&
\sim
\frac{2}{\pi R}
\int_0^1 
x^{2qR} 
\frac{1}{(4-x)\sqrt{q^2+\frac{x}{4-x}}}
\cos\left[2R\arccos\left(1-\frac{x}{2}\right)-\arctan\frac{1}{q}\sqrt{\frac{x}{4-x}}+\pi\right]
dx
\tag\egc
\endalign
$$
and 
$$
\spreadlines{3\jot}
\align
&\!\!
\int_0^1 \left(T^{(1)}(R,v';x)\right)^2 dx
\sim
\\
&\!\!
\frac{2}{\pi R}
\int_0^1 
x^{2qR} 
\frac{1}{(4-x)\sqrt{q^2+\frac{x}{4-x}}}
\left\{
1+
\cos\left[2R\arccos\left(1-\frac{x}{2}\right)-\arctan\frac{1}{q}\sqrt{\frac{x}{4-x}}+\pi\right]
\right\}
dx,
\\
\tag\egd
\endalign
$$
as $R$ and $v'$ approach infinity so that (\ega) holds.

By Lemma {\tfa}, (\egc) and (\egd) it follows that 
$$
\omega_c(2v-R,R)
\sim
\frac{16}{\pi}
\left|
\int_0^1 
x^{2qR} 
\frac{1}{(4-x)\sqrt{q^2+\frac{x}{4-x}}}
dx
\right|,
\tag\ege
$$
as $R$ and $v$ approach infinity so that $2v-R=qR$.

However, the asymptotics of the integral in (\ege) follows directly from Corollary 10.3 of \cite{\sc} (simply take $a(x)=0$ and $c(x)=0$ to account for the missing cosine factor). We obtain that
$$
\int_0^1 
x^{2qR} 
\frac{1}{(4-x)\sqrt{q^2+\frac{x}{4-x}}}
dx
\sim
\frac{1}{3q\sqrt{q^2+\frac13}}\frac{1}{R},\ \ \ R\to\infty.
\tag\egf
$$
Together with (\ege), this yields
$$
\omega_c(2v-R,R)
\sim
\frac{16}{3\pi q\sqrt{q^2+\frac13}}\frac{1}{R},
\tag\egg
$$
which proves the first equality in Theorem {\tba}.

To obtain the second equality, note that the midpoint of the top side of the gap $O_1$ in Figure {\fbe} is at distance $R$ above the horizontal dotted line in that figure, and at distance $q\sqrt{3}R$ from the vertical dotted line. It follows that the right hand side of (\egg) can be expressed in terms of the distances between the gap $O_1$ and its images $O_2$, $O_3$ and $O_4$ (see Figure {\fbe}) as
$$
\spreadlines{3\jot}
\align
\frac{16}{3\pi q\sqrt{q^2+\frac13}}\frac{1}{R}
&=
\frac{32}{\pi}\frac{2R}{(2q\sqrt{3}R)\left(2\sqrt{(q\sqrt{3}R)^2+R^2}\right)}\\
&\sim
\frac{32}{\pi}\frac{\de(O_1,O_2)}{\de(O_1,O_3)\de(O_1,O_4)}\\
&=
\frac{32}{\pi}
\sqrt{\frac{\de(O_1,O_2)\de(O_3,O_4)}{\de(O_1,O_3)\de(O_1,O_4)\de(O_2,O_3)\de(O_2,O_4)}}.
\endalign
$$
This completes the proof of Theorem {\tba}.

\mysec{8. Conjectured interaction of holes with boundary for arbitrary regions}

The example worked out in this paper adds an important new case --- the first dealing with mixed boundary conditions --- to the small collection of circumstances in which the interaction of holes with the boundary of a region (constrained or free) is determined rigorously (see \cite{\sc} for the interaction of holes with a straight line with constrained bondary, \cite{\free} for the interaction of a hole with a straight line with free boundary, and \cite{\anglep} (resp., \cite{\aanglep}) for the interaction of a gap with a $60^\circ$ (resp., $120^\circ$) angle with constrained boundary). 

Based on this we present below what we conjecture to be the interaction of gaps in a sea of dimers with the boundary in the general case.

Let $\Omega$ be a simply connected open set in the plane so that $\partial \Omega$ consists of a finite union of straight line segments (finite or infinite in length) with polar directions belonging to the set $\{0,\pm\pi/6,\pm\pi/3,\pm\pi,\pm2\pi/3,\pm5\pi/6,\pi\}$ (see the picture on the bottom in Figure {\fib} for an example). Color the boundary line segments with polar directions in the set $\{0,\pm\pi/3,\pm2\pi/3\}$ red (indicated by solid lines in Figure {\fib}), and the boundary line segments with polar directions in the set $\{\pm\pi/6,\pm5\pi/6,\pi\}$ blue (indicated by dashed lines in Figure {\fib}). Let $a_1,\dotsc,a_k$ be distinct points in the interior of $\Omega$.

\topinsert
\twoline{\mypic{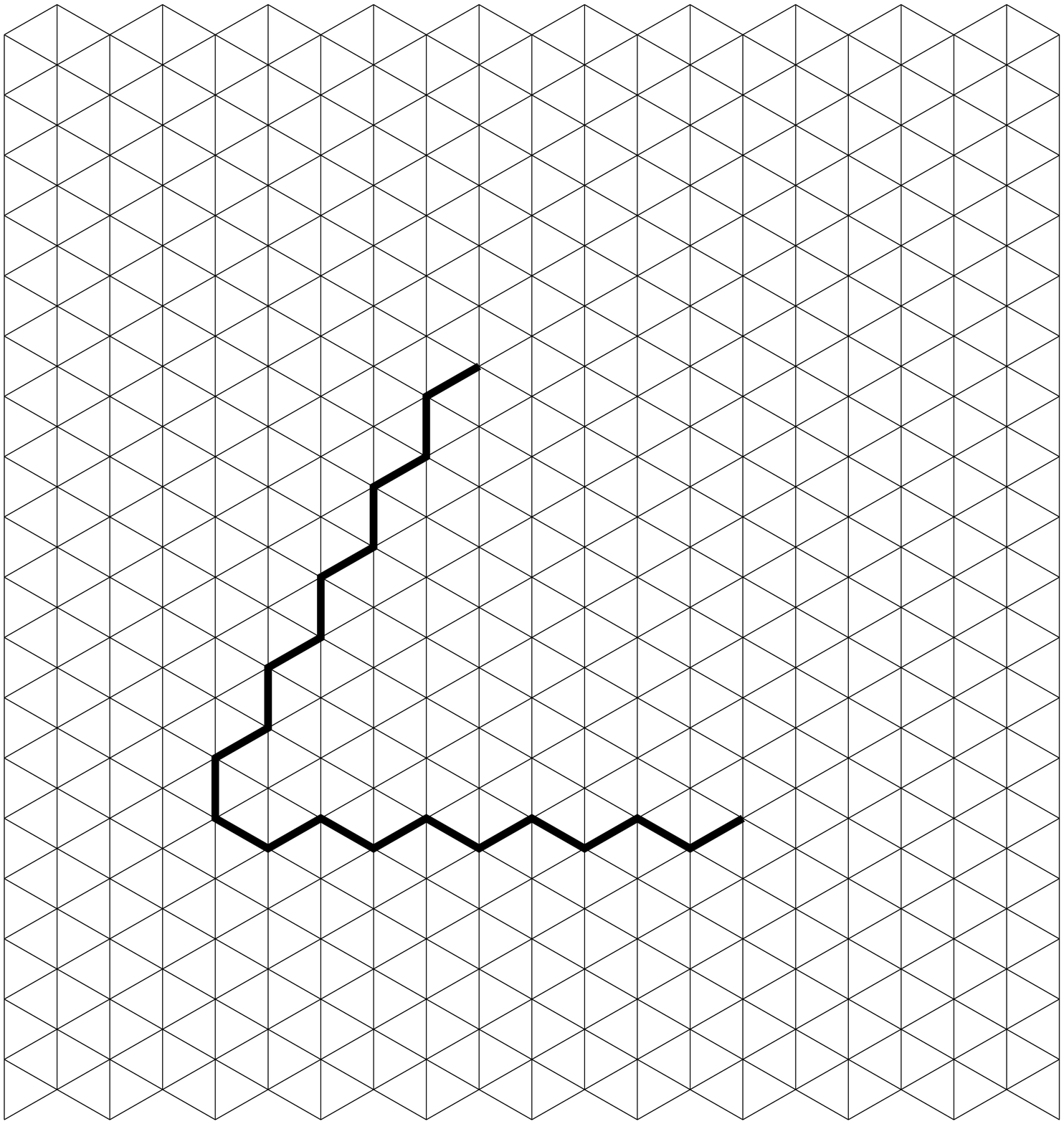}}{\mypic{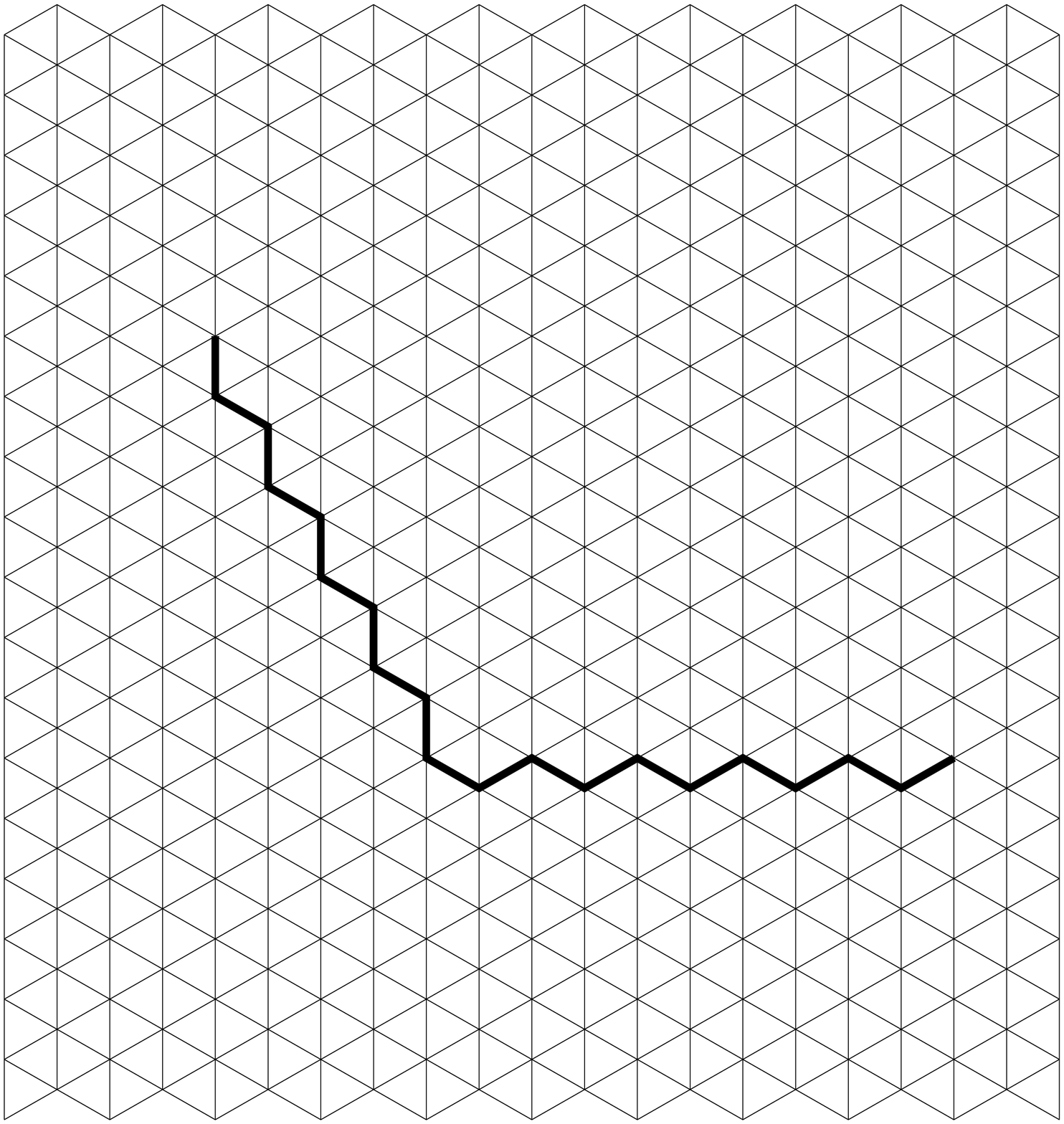}}
\centerline{Figure~{\fia}. {\rm The two types of zig-zag corners in $\Omega_n$.}}
\endinsert

For $n\geq1$, let $\Omega_n$ be a lattice region on the triangular lattice (drawn so that one family of lattice lines is vertical) whose boundary is the union of ``straight zig-zag'' line segments (each picture in Figure {\fia} shows two such line segments) and lattice line segments (see the picture on top in Figure {\fib} for an example). We assume that each corner where two zig-zag portions of the boundary meet looks, up to rotation by some multiple of $\pi/3$, as shown in Figure {\fia}.
Let $\Omega_n$ have constrained boundary conditions along the zig-zag portions, and free boundary conditions along the lattice segment portions. Suppose that in the scaling limit\footnote{ By the scaling limit we mean here the fine mesh limit, i.e. the limit as the lattice spacing of the triangular lattice on which the regions $\Omega_n$ reside approaches zero.}we have $\Omega_n\to\Omega$ as $n\to\infty$, in such a way that the zig-zag portions of the boundary of $\Omega_n$ approach red segments on the boundary of $\Omega$, and the free portions of the boundary of $\Omega_n$ approach blue segments on the boundary of $\Omega$.

Let $O_1^{(n)},\dotsc,O_k^{(n)}$  be finite unions of unit triangles from the interior of $\Omega_n$, so that for any fixed $i$, the $O_i^{(n)}$'s are translates of one another for all $n\geq1$ (these will be the gaps). Assume that in the scaling limit as $n\to\infty$, $O_i^{(n)}$ shrinks to $a_i$, for $i=1,\dotsc,k$. Define $\q(O_i^{(n)})$ to be equal to the number of right-pointing unit triangles minus the number of left-pointing unit triangles in $O_i^{(n)}$.

\topinsert
\centerline{\mypic{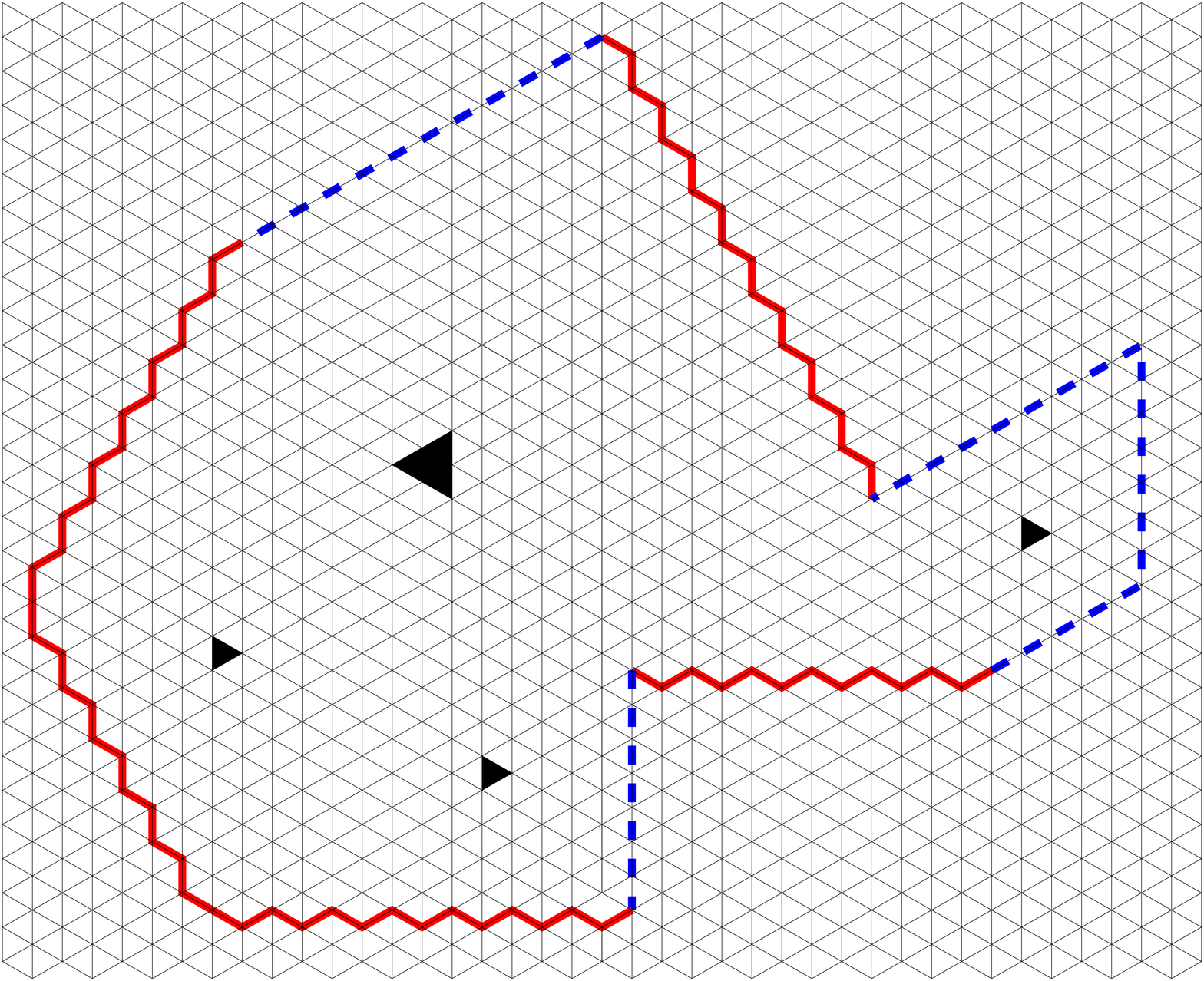}}
\bigskip
\centerline{\mypic{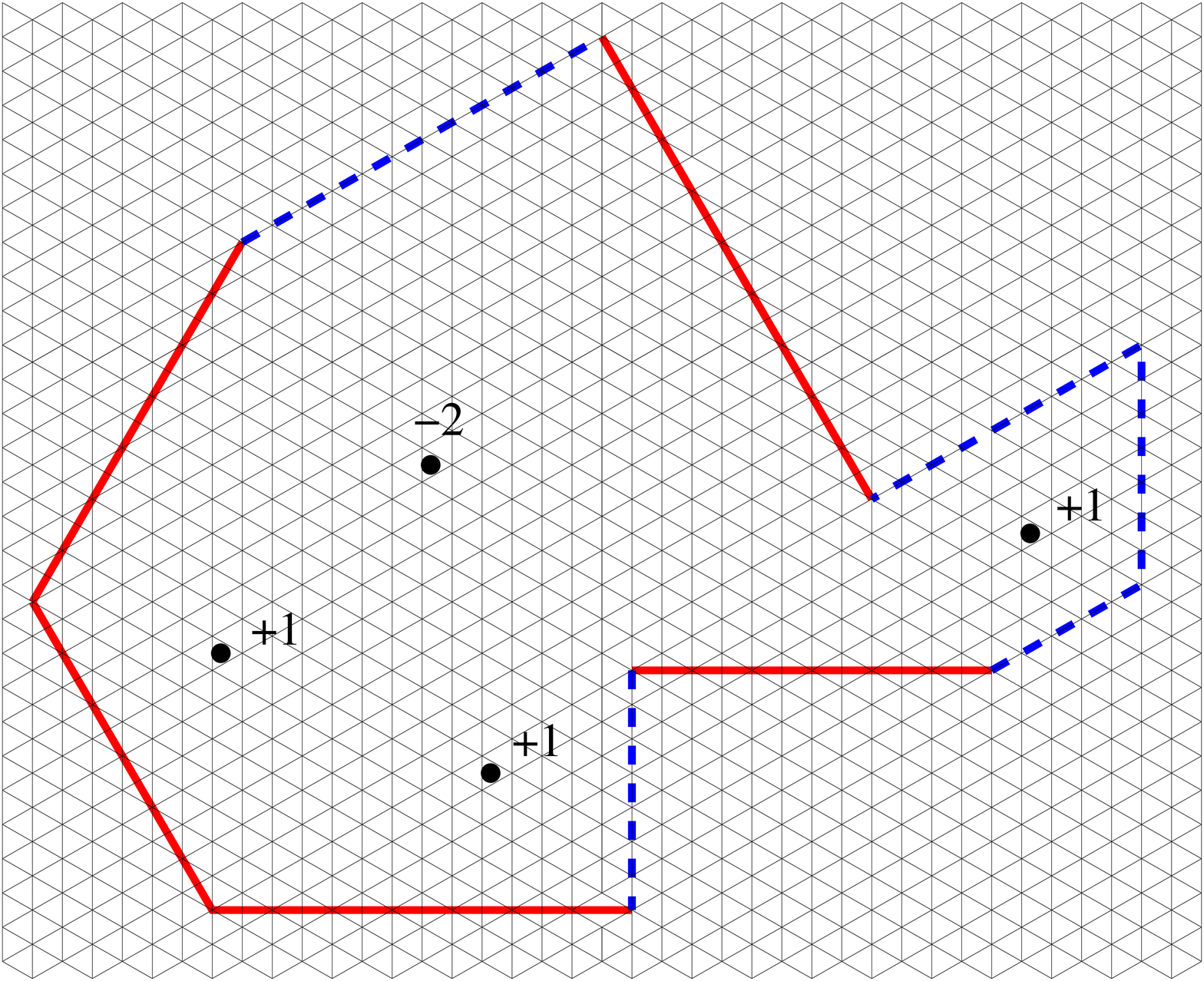}}
\centerline{{\smc Figure~{\fib}. {\rm Gaps interacting with the boundary of a general region,}}}
\centerline{{{\rm \ \ \ \ and the corresponding steady state heat flow problem.}}}
\endinsert

Corresponding to these regions, consider a 2D physical system $S$ consisting of a uniform block of material having shape $\bar\Omega$, with $k$ point sources at $a_1,\dotsc,a_k$, of strengths given by the $\q(O_i^{(n)})$'s (a heat source with a negative strength is a heat sink; see the picture on the bottom of Figure {\fib}). Let the boundary of this block of material be a perfect insulator along its red segments (which correspond to the constrained boundaries of $\Omega_n$), and let it be kept at a common constant temperature along its blue segments (corresponding to the free boundaries of $\Omega_n$). Denote by $E$ the heat energy\footnote{ A good way to think about the heat energy is that, up to a multiplicative constant (which cancels out when taking the ration on the right hand side of (\eia)), it equals $\int_\Omega h^2$, where $h$ is the magnitude of the heat flow vector.} of this system in the steady state.

Then we believe that the behavior of the correlation of the gaps with the boundary is given in the scaling limit by the following.

\proclaim{Conjecture \tia} Let ${O'_i}^{(n)}$'s be translations of the $O_i^{(n)}$'s that shrink to distict points $a'_1,\dotsc,a'_k\in\Omega$ in the scaling limit as $n\to\infty$. Then
$$
\frac{\M_f(\Omega_n\setminus O_1^{(n)}\cup\cdots\cup O_k^{(n)})}
{\M_f(\Omega_n\setminus {O'_1}^{(n)}\cup\cdots\cup {O_k'}^{(n)})}
\to
\frac{\exp(-E)}{\exp(-E')},\tag\eia
$$
where $E'$ is the heat energy of the system obtained from $S$ by moving the point heat sources to positions $a'_1,\dotsc,a'_k$.

\endproclaim

We note that all the five examples of special circumstances that have been worked out in the literature (which are mentioned in the first paragraph of this section) conform to this conjecture. This can be seen by making the following two observations. On the one hand, the interaction of the holes with the boundary is given in all these four cases by a signed analog of the method of images in electrostatics, in which the mirror image of a gap (which is the analog of a charge in our context) has opposite orientation (and thus $q$-value of opposite sign) if the portion of the boundary in which the reflection occurs is free, but has the same orientation (and hence the same $q$-value) as the original gap if the portion of the boundary in which the reflection occurs is constrained. On the other hand, the corresponding 2D steady state heat flow problem can also be solved by this same signed method of images, noticing that straight line perfect insulator boundary portions can be simulated by taking image heat sources of the same sign, and straight line constant temperature boundary portions by taking image heat sources of the opposite sign.

This shows that from the point of view of the interaction of gaps with boundaries in a sea of dimers, the above heat interpretation is more natural than the electrostatic interpretation (to which it is equivalent in the bulk, see \cite{\sc}) that has been highligthed in our previous work \cite{\sc}\cite{\ec}\cite{\ef}\cite{\ov}. This is because there is no circumstance in electrostatics that leads to taking image charges of the same sign, as this would correspond to a straight line conductor separating two media whose permittivities have infinite ratio, which is not possible because the permittivities are always finite and positive (see e.g. \cite{\Feyntwo, \S12}).





%
%

\mysec{9. Concluding remarks}

It is a striking fact that the number $SSC(a,b,c)$ of symmetric, self-complementary plane partitions that fit in a $2a\times2b\times2c$ box is equal to the total number $P(a,b,c)$ of plane partitions that fit in an $a\times b\times c$ box (see e.g. \cite{\Sta}):
$$
SSC(2a,2b,2c)=P(a,b,c).\tag\eja
$$
It would be interesting to find an analog of (\eja) in the context of our generalization of the symmetric, self-complementary symmetry class given in Corollary {\tcd}.

We end this paper by noting that the interaction of $k$ left-pointing holes of side two with the 90 degree mixed boundary angle can be proved using the same approach we presented here, using (\ede) as starting point. To carry it out, we need an extension of the above calculations in the style of the special case of the situation treated in \cite{\sc} when all holes have the same orientation, and there is no unit~hole. The resulting interaction conforms again to Conjecture {\tia}.

\mysec{References}
{\openup 1\jot \frenchspacing\raggedbottom
\roster

\myref{\And}
  G. E. Andrews, Plane partitions (III): The weak Macdonald
conjecture, {\it Invent. Math.} {\bf 53} (1979), 193--225.

\myref{\ri}
  M. Ciucu, Rotational invariance of quadromer correlations on the hexagonal lattice, \
{\it Adv. in Math.} {\bf 191} (2005), 46-77.

\myref{\sc}
  M. Ciucu, A random tiling model for two dimensional electrostatics, {\it Mem. Amer. \
Math. Soc.} {\bf 178} (2005), no. 839, 1--106.

\myref{\ec}
  M. Ciucu, The scaling limit of the correlation of holes on the triangular lattice
with periodic boundary conditions, {\it Mem. Amer. Math. Soc.} {\bf 199} (2009),
no. 935, 1-100.

\myref{\ef}
  M. Ciucu, The emergence of the electrostatic field as a Feynman sum in random
tilings with holes, {\it Trans. Amer. Math. Soc.} {\bf 362} (2010), 4921-4954.

\myref{\ov}
  M. Ciucu, Dimer packings with gaps and electrostatics, {\it Proc. Natl. Acad. Sci.
USA} {\bf 105} (2008), 2766-2772.

\myref{\free}
  M. Ciucu and C. Krattenthaler, The interaction of a gap with a free boundary in a
two dimensional dimer system, {\it Comm. Math. Phys.} {\bf 302} (2011), 253-289.

\myref{\gd}
  M. Ciucu, The interaction of collinear gaps of arbitrary charge in a two
dimensional dimer system, 2012, arXiv:1202.1188.

\myref{\anglep}
  M. Ciucu and I. Fischer, A triangular gap of side 2 in a sea of dimers in a $60^\circ$ angle, {\it J. Phys. A: Math. Theor.} {\bf 45} (2012), 494011.

\myref{\aanglep}
  M. Ciucu, A triangular gap of side 2 in a sea of dimers in a $120^\circ$ angle, preprint, 2013.

\myref{\fakt}
  M. Ciucu and C. Krattenthaler, A factorization theorem for rhombus tilings of a hexagon with triangular holes, preprint, 2013.

\myref{\DT}
  G. David and C. Tomei, The problem of the calissons,
{\it Amer\. Math\. Monthly} {\bf 96} (1989), 429--431.

\myref{\Feyntwo}
   R. P. Feynman, ``The Feynman Lectures on Physics,'' vol. II, Addison-Wesley, Reading, Massachusetts, 1964.

\myref{\FS}
  M. E. Fisher and J. Stephenson, Statistical mechanics of dimers on a plane
lattice. II. Dimer correlations and monomers, {\it Phys. Rev. (2)} {\bf 132} (1963),
1411--1431.

\myref{\GV}
  I. M. Gessel and X. Viennot, Binomial determinants, paths, and hook length formulae,
{\it Adv. in Math.} {\bf 58} (1985), 300--321.

\myref{\Kuo}
Kuo, Eric H. Applications of graphical condensation for enumerating matchings and tilings. Theoret. Comput. Sci. 319 (2004), no. 1-3, 29–57.

\myref{\Kup}
  G. Kuperberg, Symmetries of plane partitions and the permanent-de\-ter\-mi\-nant
method, {\it J. Combin. Theory Ser. A} {\bf 68} (1994), 115--151.

\myref{\Lind}
  B. Lindstr\"om, On the vector representations of induced
matroids, {\it Bull. London Math. Soc.} {\bf 5} (1973), 85--90.

\myref{\Proc}
  Proctor, Odd symplectic groups, {\it Invent. Math.} {\bf 92} (1988), 307--332.

\myref{\Schur}
  I. Schur, \"Uber die Darstellung der symmetrischen und der alternierenden Gruppe durch gebrochene lineare Substitutionen, {\it J. Reine Angew. Math.} {\bf 139} (1911), 155-250. 

\myref{\Sta}
  R. P. Stanley, Symmetries of plane partitions, {\it J. Comb. Theory Ser. A} {\bf 43} (1986), 103--113.

\myref{\Ste}
  J. R. Stembridge, Nonintersecting paths, pfaffians and plane partitions,
{\it Adv. in Math.} {\bf 83} (1990), 96--131.

\endroster\par}

\enddocument